\documentclass{emulateapj}

\usepackage{graphicx} 
\usepackage{longtable} 
\usepackage{rotating}
\usepackage{tabularx}

\usepackage{array} 
\usepackage{amssymb,amsmath} 

\newcommand{\numax}{\mbox{$\nu_{\rm max}$}}

\newcommand{\kep}{\mbox{\textit{Kepler}}}
\newcommand{\teff}{\mbox{$T_{\rm eff}$}}
\newcommand{\logg}{\mbox{$\log g$}}
\newcommand{\feh}{\mbox{$\rm{[Fe/H]}$}}
\newcommand{\msun}{\mbox{$M_{\sun}$}}
\newcommand{\rsun}{\mbox{$R_{\sun}$}}

\newcommand{\nstars}{\mbox{196,468}}
\newcommand{\nuncl}{\mbox{11,532}}
\newcommand{\nast}{\mbox{2,762}}

\include{journals}

\slugcomment{accepted for publication in ApJS}

\shorttitle{Revised Stellar Properties of Kepler Targets}
\shortauthors{Huber et al.}

\begin{document}

\title{Revised Stellar Properties of Kepler Targets for the \\ 
Quarter 1--16 Transit Detection Run}

\author{
Daniel Huber\altaffilmark{1,2}, 
Victor Silva Aguirre\altaffilmark{3}, 
Jaymie M.~Matthews\altaffilmark{4}, 
Marc H.~Pinsonneault\altaffilmark{5}, 
Eric Gaidos\altaffilmark{6}, 
Rafael A.~Garc\'\i a\altaffilmark{7}, 
Saskia Hekker\altaffilmark{8,9}, 
Savita Mathur\altaffilmark{10}, 
Beno\^it Mosser\altaffilmark{11}, 
Guillermo Torres\altaffilmark{12}, 
Fabienne A.~Bastien\altaffilmark{13}, 
Sarbani Basu\altaffilmark{14}, 
Timothy R.~Bedding\altaffilmark{15,3}, 
William J.~Chaplin\altaffilmark{16,3}, 
Brice-Olivier Demory\altaffilmark{17}, 
Scott W.~Fleming\altaffilmark{18}, 
Zhao Guo\altaffilmark{19}, 
Andrew W.~Mann\altaffilmark{20}, 
Jason F.~Rowe\altaffilmark{1,2}, 
Aldo M.~Serenelli\altaffilmark{21}, 
Myron A.~Smith\altaffilmark{22}, and 
Dennis Stello\altaffilmark{15,3} 
}
\altaffiltext{1}{NASA Ames Research Center, Moffett Field, CA 94035, USA; daniel.huber@nasa.gov}
\altaffiltext{2}{SETI Institute, 189 Bernardo Avenue, Mountain View, CA 94043, USA}
\altaffiltext{3}{Stellar Astrophysics Centre, Department of Physics and Astronomy, Aarhus University, Ny Munkegade 120, DK-8000 Aarhus C, Denmark}
\altaffiltext{4}{Department of Physics and Astronomy, University of British Columbia, Vancouver, Canada}
\altaffiltext{5}{Department of Astronomy, Ohio State University, OH 43210, USA}
\altaffiltext{6}{Department of Geology and Geophysics, University of Hawaii at Manoa, Honolulu, HI 96822, USA}
\altaffiltext{7}{Laboratoire AIM, CEA/DSM-CNRS, Universit\'e Paris 7 Diderot, IRFU/SAp, Centre de Saclay, 91191, Gif-sur-Yvette, France}
\altaffiltext{8}{Max-Planck-Institut f\"ur Sonnensystemforschung, Justus-von-Liebig-Weg 3, 37077 G\"ottingen, Germany}
\altaffiltext{9}{Astronomical Institute 'Anton Pannekoek', University of Amsterdam, Science Park 904, 1098 XH Amsterdam, The Netherlands}
\altaffiltext{10}{Space Science Institute, 4750 Walnut Street Suite 205, Boulder, CO 80301, USA}
\altaffiltext{11}{LESIA, CNRS, Universit\'e Pierre et Marie Curie, Universit\'e Denis, Diderot, Observatoire de Paris, 92195 Meudon cedex, France}
\altaffiltext{12}{Harvard-Smithsonian Center for Astrophysics, Cambridge, Massachusetts 02138, USA}
\altaffiltext{13}{Department of Physics \& Astronomy, Vanderbilt University, 1807 Station B, Nashville, TN 37235, USA}
\altaffiltext{14}{Department of Astronomy, Yale University, New Haven, CT 06511, USA}
\altaffiltext{15}{Sydney Institute for Astronomy (SIfA), School of Physics, University of Sydney, NSW 2006, Australia}
\altaffiltext{16}{School of Physics and Astronomy, University of Birmingham, Birmingham B15 2TT, UK}
\altaffiltext{17}{Department of Physics, Massachusetts Institute of Technology, 77 Massachusetts Ave., Cambridge, MA 02139, USA}
\altaffiltext{18}{Space Telescope Science Institute, 3700 San Martin Drive, Baltimore, Maryland 21218, USA}
\altaffiltext{19}{Center for High Angular Resolution Astronomy, Georgia State University, PO Box 3969, Atlanta, GA 30302, USA}
\altaffiltext{20}{Institute for Astronomy, University of Hawaii, 2680 Woodlawn Drive, Honolulu, HI 96822, USA}
\altaffiltext{21}{Instituto de Ciencias del Espacio (CSIC-IEEC), Facultad de Ciencias, 08193 Bellaterra, Spain}
\altaffiltext{22}{NOAO, 950 N. Cherry Ave., Tucson, AZ 85719, USA}

\begin{abstract}
We present revised properties for 196,468 stars observed by the NASA Kepler Mission and
used in the analysis of Quarter 1--16 (Q1--Q16) data to detect and characterize transiting planets.
The catalog is based on a compilation of 
literature values for atmospheric properties (temperature, surface gravity, and metallicity) 
derived from different observational techniques (photometry, spectroscopy, 
asteroseismology, and exoplanet transits), which were then 
homogeneously fitted to a grid of Dartmouth stellar isochrones. 
We use broadband photometry and asteroseismology 
to characterize \nuncl\ \kep\ targets which were previously unclassified in the 
Kepler Input Catalog (KIC). We report the detection of oscillations in \nast\
of these targets, classifying them as giant stars and increasing the 
number of known oscillating giant stars observed by \kep\ by $\sim20\%$ to a total of 
$\sim$\,15,500 stars. Typical uncertainties in derived radii and masses  
are $\sim 40$\% and $\sim 20$\%, respectively, for stars with photometric constraints only, 
and $5-15$\% and $\sim 10$\% for stars based on spectroscopy and/or asteroseismology, 
although these uncertainties vary strongly with spectral type and luminosity class.
A comparison with the Q1--Q12 catalog shows a systematic decrease in the
radii of M dwarfs, while radii for K dwarfs decrease or increase depending on the Q1--Q12 
provenance (KIC or Yonsei-Yale isochrones). Radii of F--G dwarfs are on average 
unchanged, with the exception of newly identified giants. 
The Q1--Q16 star properties catalog is a first step towards an improved characterization of 
all \kep\ targets to support planet occurrence studies.
\end{abstract}


\section{Introduction}
The unprecedented precision of photometric data collected by the 
NASA \kep\ mission \citep{borucki10,koch10b} has revolutionized 
planetary and stellar astrophysics over the past years. 
Examples of breakthrough discoveries in exoplanet science include more than 2700 new
planet candidates \citep{borucki11b,borucki11,batalha12,burke14}, 
measurements of planet densities in multi-planet systems through transit timing 
variations \citep{holman10,lissauer11,carter12}, the detection of 
small planets in or near the habitable zone \citep{borucki12,borucki13,barclay13}, 
and the discovery of single and multi-planet systems around binary stars 
\citep{doyle11,welsh12,orosz12}. At the same time, \kep\ data allowed key 
advances in stellar astrophysics such as the detection of 
more than 2000 eclipsing binary stars \citep{prsa11,slawson11,matijevic12} including eclipsing 
triple systems \citep{carter11,derekas11}, 
the study of the core structure and rotation of subgiant and 
red-giant stars \citep{bedding11,beck11,mosser12,deheuvels12}, 
an order of magnitude increase of known
dwarf stars with detected oscillations \citep{chaplin11a}, 
and the discovery of tidal pulsations in eccentric binary systems \citep{thompson12}.

In addition to characterizing individual exoplanet systems, 
a primary mission goal of \kep\ is to determine the frequency of Earth-sized 
planets in the habitable zones of Sun-like stars. 
Planet occurrence rates crucially depend not only on our knowledge 
of properties such as radii and luminosities 
of the host stars (which in turn determine the properties of the planets), 
but also on our understanding of the properties of the 
\textit{parent sample}. For example, if a significant 
number of subgiant or giant stars have been misclassified as dwarfs, this 
would bias occurrence 
rates since small planets are harder to detect around larger stars. 

Many studies which have explored planet-occurrence rates using the \kep\ sample
\citep[e.g.,][]{catanzarite11,howard11,traub12,dong13b,fressin13,petigura13,petigura13b} 
have mostly relied on stellar properties based on 
the Kepler Input Catalog \citep[KIC,][hereafter B11]{brown11}. 
The primary purpose of the KIC was to discern dwarfs from giants to optimize 
the target selection towards finding Earth-sized planets in the habitable zones 
of Sun-like stars
\citep{batalha10}. As emphasized by B11, the methodology behind constructing 
the KIC limits its use beyond target selection. 
Recent planet-occurrence studies have used improved stellar properties 
either for specific parameter ranges such as cool dwarfs \citep{dressing13,morton13},  
or for a wide range of spectral types for dwarfs \citep{gaidos13b}. However, a 
revised characterization of the full target sample to support the \kep\ planet-detection 
pipeline and planet-occurrence studies has yet to be completed.

Since the creation of the KIC, a large number of new observations 
have become available. For example, 
new broadband photometry covering the full \kep\ field has been completed 
\citep{everett12,greiss12}. 
Additionally, a large amount of spectroscopic follow-up observations have been performed 
within the Kepler Community Follow-Up 
Program (CFOP\footnote{https://cfop.ipac.caltech.edu/home/login.php}), and 
systematic spectroscopic surveys of the \kep\ field using multi-object fiber-fed 
spectrographs are currently in progress \citep{zasowski13,apokasc}. Importantly, \kep\ 
light curves themselves contain information about fundamental properties of stars. 
In particular, \kep\ has allowed the application of asteroseismology to stars ranging from 
hot, compact objects \citep{kawaler10,ostensen11}, 
to classical pulsators \citep{grigahcene10,kolenberg10,kurtz11}, 
cool dwarfs \citep{chaplin11a,silvaaguirre11,mathur12} and red giant stars 
\citep{bedding10b,hekker11c,kallinger10,mosser11c}. 

Many of these results have not yet 
been taken into account in planet occurrence studies due to the lack of
catalogs covering most \kep\ targets. 
In this paper we present a catalog of revised 
properties for \nstars\ \kep\ targets based on a consolidation of 
literature values and the first
characterization of unclassified stars in the KIC.

\section{Kepler Input Catalog}
\label{sec:kic}

We begin with a brief review of the Kepler Input Catalog. As described by 
B11, the primary observables for the KIC stellar classification 
pipeline (SCP) were KIC $griz$ and 2MASS $JHK$ broadband photometry \citep{skrutskie06}, 
supplemented by an intermediate-band $D51$ filter (centered 
on the Mg Ib lines at 510\,nm). These data were used to calculate seven independent colors, which 
were then compared to synthetic colors calculated from ATLAS9 model atmospheres \citep{castelli04}. 
To account for interstellar extinction, B11 adopted a simple reddening model giving 1 
magnitude of $V$-band extinction per 1\,kpc in the galactic plane, 
decreasing with galactic latitude with an $e$-folding scale height of 150\,pc.
To overcome degeneracies of matching broadband colors to models 
to estimate effective temperature (\teff), surface gravity (\logg) and metallicity (\feh), 
three priors were adopted: a metallicity prior based on a solar-neighborhood 
distribution of the Geneva-Copenhagen survey \citep{nordstroem}, a \teff-\logg\
prior based on the number density of stars in the Hipparcos catalog \citep{vanleeuwen07b}, 
and a prior on the number density of stars as a function of galactic 
latitude. Stellar masses and luminosities were derived from 
average relations between \teff, \logg, luminosity and mass calculated from 
Padova isochrones \citep{girardi00}. Finally, stellar radii were calculated from the 
derived effective temperatures and luminosities.

The difficulty of estimating stellar properties (in particular 
\logg\ and \feh) 
from broadband colors resulted in a number of shortcomings which 
limited the use of the KIC beyond target selection. Follow-up 
studies have since tested stellar properties in the KIC to quantify these 
shortcomings. The main conclusions of these tests can be summarized as 
follows:

\begin{itemize}
\item KIC surface gravities
are frequently overestimated by up to 0.2\,dex, resulting in 
underestimated radii by up to 50\% \citep{verner11,everett13}. 
Observational biases also suggest that the fraction of subgiant stars in the 
\kep\ target sample may be underestimated \citep{gaidos13}.

\item KIC temperatures are on average 200\,K cooler than temperatures based on the 
Sloan system or the infrared flux method \citep[][hereafter P12]{pinsonneault11}. 
KIC $griz$ photometry also 
shows a color-dependent offset to Sloan DR9 photometry.

\item A large fraction 
of bright ($Kp<14$) late-K to mid-M type stars are misclassified as dwarfs \citep{mann12}.
Furthermore, surface gravities, metallicities and radii for genuine late-type dwarfs 
are systematically biased \citep{muirhead12,batalha12,dressing13}.

\item Roughly 5\% of Kepler targets have KIC parameters that 
should be absent in a well-studied field population, specifically
G-type dwarfs with $\logg \sim 5$ and K-dwarfs with 
$\teff \sim 5000$\,K and $\logg \sim 4.2$ \citep{batalha12}. We refer to these stars 
in the following as ``No-Man's-Land'' stars.
\end{itemize}

Similar conclusions have been found theoretically by \citet{farmer13}, 
who processed a synthetic stellar population of the \kep\ field 
through the KIC classification pipeline. 
We note that most of the above shortcomings have already been anticipated and 
emphasized by B11.

\section{Stellar Models}
\label{sec:models}

In general, determining stellar masses and radii
involves a comparison of observations with models. For the 
current catalog, we adopted the 2012 isochrones from the Dartmouth Stellar Evolution Database 
\citep[DSEP,][]{dotter08}\footnote{http://stellar.dartmouth.edu/models/index.html}. 
We have used the DSEP interpolation routine to 
produce a grid of $1-15$\,Gyr isochrones in steps of 0.5\,Gyr in age and 0.02\,dex in 
\feh. Only models with solar-scaled alpha-element abundances have been included. 
In addition to low-mass models, we supplemented the grid with $0.25-1$\,Gyr isochrones with 
the default metallicity spacing  ($-2.5$, $-2.0$, $-1.5$, $-1.0$, $-0.5$, 0.07, 
0.15, 0.36 and 0.5 dex). The full grid includes approximately $1.5\times10^6$ individual models.

The choice of DSEP was motivated by 
the good agreement with models by \citet{baraffe}, which have been 
demonstrated to reasonably reproduce observations of low-mass dwarfs 
with empirically measured radii, masses and effective temperatures from 
long-baseline interferometry \citep{boyajian12b,boyajian13} or low-mass eclipsing 
binary systems \citep{kraus11,carter11} 
\citep[although significant offsets still exist, see][]{boyajian13}.
Additionally, DSEP models cover a large parameter space and 
include broadband colors based on PHOENIX model atmospheres \citep{hauschildt99} 
for most filters with available data for \kep\ targets. We note that the 
adopted isochrone grid does not include He-core burning models for stars which undergo 
the helium flash ($M\lesssim2\,\msun$). As emphasized in Section \ref{sec:shortcomings}, 
this introduces a significant 
bias for derived masses and radii of giant stars in the catalog. 

The coolest DSEP models for dwarfs have temperatures close to 3200\,K. 
However, recent spectroscopic follow-up observations revealed 
a significant number of ultra-cool dwarfs in the Kepler field \citep{martin13}, and additional 
late-type M dwarfs have been added to the \kep\ target list through 
Guest Observer programs\footnote{http://keplerscience.arc.nasa.gov/}. 
To characterize these stars, we  
fit second to fourth order polynomials between colors, temperatures, gravities, 
and radii to $>2$\,Gyr BT-Settl isochrones \citep{allard12} with temperatures 
between $2000-3400$\,K. 
These polynomials were then used to interpolate the BT-Settl grid and provide 
typical stellar properties for a given color or temperature.

\section{Consolidation of Literature Values}
\label{sec:consol}

We have collected published values for 
\teff, \logg\ and \feh\ for all \kep\ targets. 
We  only considered publications that have derived stellar properties for more than one 
star, with the exception of confirmed exoplanet host stars.
We considered five main data sources:

\begin{itemize}
\item Asteroseismology: Stellar oscillations 
provide accurate measurements of stellar properties such as
density and surface gravity 
\citep{stello08,bedding11b,miglio12b,huber12b,silva12,hekker13}. For cool stars with 
asteroseismic detections but without published $\logg$ values, we have used the 
observed frequency of maximum power (\numax) to 
estimate \logg\ through the scaling relation $\numax \propto g\,\teff^{-0.5}$ 
\citep{brown91,belkacem11}, 
where \teff\ was adopted either from photometry or spectroscopy, as described below.

\item Transits: Transiting exoplanets 
allow accurate measurements of the 
mean stellar density if the orbital eccentricity and impact parameter are accurately known 
\citep[see, e.g.,][]{seager03,brown10,winn10b}.
For multi-planet systems in particular, transit-derived stellar 
densities reach uncertainties comparable 
to asteroseismology and are often preferred over constraints on \logg\ from 
spectroscopy \citep{lissauer13,jontof13}. This category also includes 
host stars in eclipsing double-lined spectroscopic binaries with
dynamically measured masses and radii, which can be combined to calculate \logg.

\item Spectroscopy: Modeling spectra is one of the most traditional methods 
to derive \teff, \logg\ and \feh, although some limitations of spectroscopic 
surface gravities exist \citep{torres12,huber13}. We have only adopted published solutions 
based on high-resolution ($R\gtrsim20000$) spectra. We note that the 
properties are based on different spectroscopic analysis pipelines 
such as SME \citep[Spectroscopy Made Easy,][]{valenti96}, 
SPC \citep[Stellar Parameter Classification,][]{buchhave12},
VWA \citep[Versatile Wavelength Analysis,][]{bruntt10}, and 
ROTFIT \citep{frasca03}, and hence are not homogeneous.

\item Photometry: Broadband photometry is a well established method to determine
temperatures \citep[e.g.,][]{casagrande10}, with narrow-band filters allowing some sensitivity to 
\logg\ and \feh. This category includes literature 
using revised KIC photometry, in particular the 
temperature scale revision by P12.

\item KIC: For this category, original 
Kepler Input Catalog values have been adopted. 
Note that this category is listed separately from photometry 
to discern it from literature values published after the launch of the \kep\ mission.
\end{itemize}

\begin{figure*}
\begin{center}
\resizebox{\hsize}{!}{\includegraphics{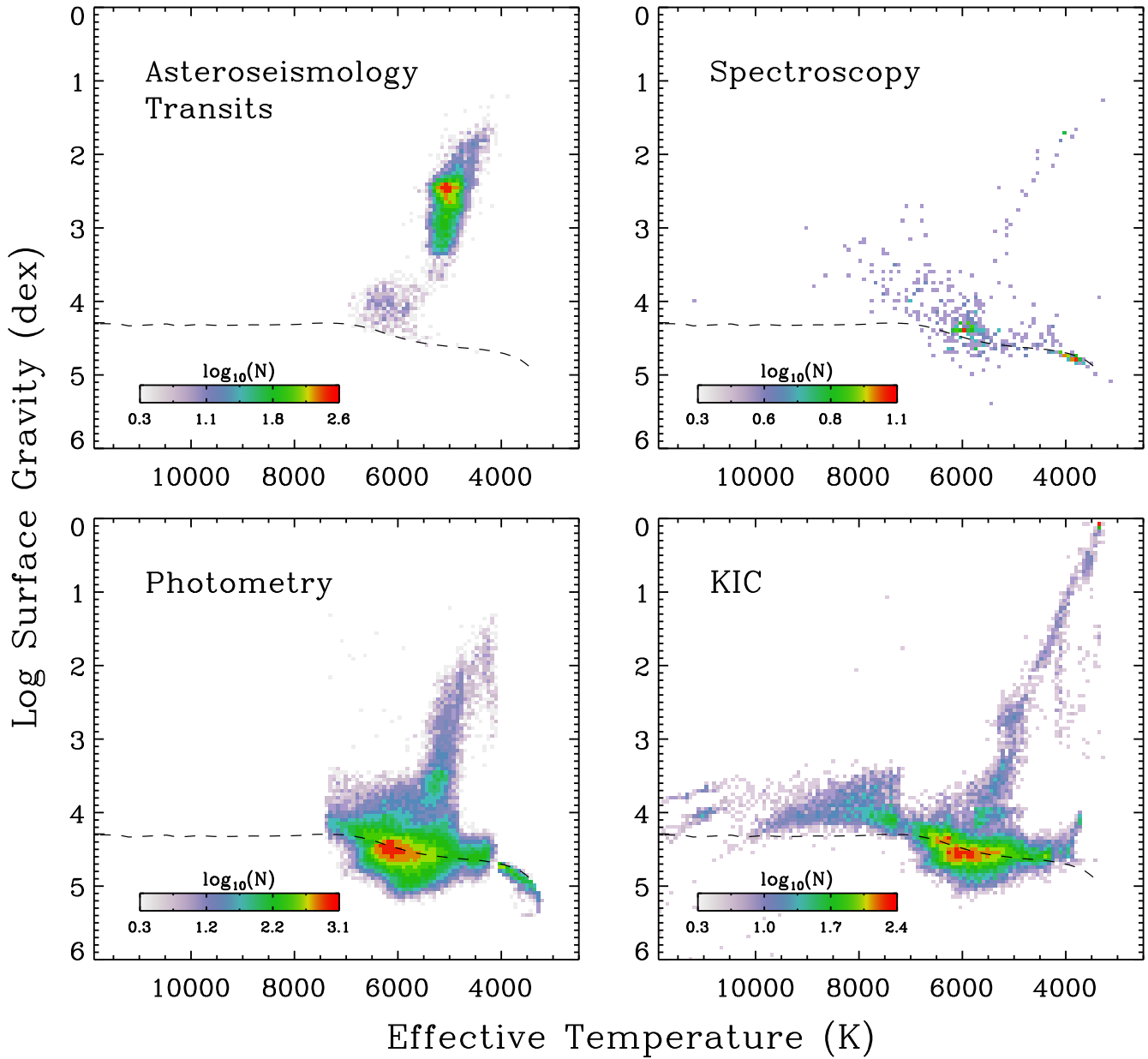}}
\caption{Surface gravity versus effective temperature for the main categories of 
consolidated literature input for the Q1--Q16 catalog. Colors denote the relative logarithmic 
number density of stars as given in the legend. The black dashed line shows the 
solar-metallicity zero-age main sequence for Dartmouth models.
Following the 
notation in Table \ref{tab:priority}, the panels include the following 
categories: $1-5$ (top left), $7-10$ (top right), $11-12$ (bottom left) and 
13 (bottom right). Note that 
unclassified stars (categories 6 and 14) are not included in this figure.}
\label{fig:categories}
\end{center}
\end{figure*}

Table \ref{tab:priority} lists the adopted prioritization scheme for stars with 
literature values from more than one source, and
Figure \ref{fig:categories} shows the main categories in a $\logg$ versus $\teff$ diagram.

Categories 1--2 include the ``gold-standard'' sample 
of $\sim$250 stars for which both high-resolution spectroscopy and either
asteroseismology or transit-derived densities are available. This combination removes 
degeneracies in the spectroscopic analysis and typically leads 
to best possible characterization of \kep\ targets (except for bright stars with 
measured parallaxes).

Categories 3--6 contain stars with asteroseismic surface gravities for which 
no spectroscopic effective temperatures or metallicities are available. 
For most of these stars, temperatures from P12 were adopted. 
To ensure consistency between the P12 temperatures (which 
were calibrated to $\feh=-0.2$) and the adopted metallicities (mostly based on KIC values), 
we have corrected the temperatures using Table 4 in 
P12\footnote{Note that due to an error in copying the table, the corrected 
temperatures are cooler than when using the original P12 corrections. 
However, the effect is only a few degrees on average and 20\,K at maximum, and hence 
negligible compared to the uncertainties.}. We emphasize 
that KIC metallicities are valid only in a statistical sense, but are not 
accurate on a star-by-star basis \citep{brown11,bruntt11,dong13}. 
Category 6 includes giant stars which were so far unclassified in the KIC, but yielded an 
asteroseismic detection in this study (see Section \ref{sec:uncl}).

Category 7 comprises stars characterized by high-resolution 
spectroscopy only. This includes large spectroscopic surveys such as \citet{buchhave12} 
for F--K dwarfs, \citet{muirhead12} for M dwarfs, and \citet{uytterhoeven11} for A--F stars.
Categories 8--10 include stars for which at least one property has been determined 
through spectroscopy, with the remaining properties constrained by photometry 
or the KIC.

Category 11 contains stars whose properties are solely based on new broadband photometry. 
This 
category is dominated by the recent revision of M-dwarf parameters by \citet{dressing13}, 
and the study of planet-candidate hosts by \citet{gaidos13b}.

Category 12 comprises stars with revised temperatures from P12 combined with 
KIC \logg\ and \feh. The same \teff\ corrections as described for Category 5 
have been adopted. We note that this category includes nearly 70\% of the full sample. 

Category 13 includes stars that only have KIC parameters available. Stars in this category
make up roughly 15\% of the target sample, and either fell 
outside the temperature range of the P12 calibration, or were not assigned 
\teff\ values due to a lack of good 2MASS photometry. To ensure consistency 
with the remaining sample, KIC temperatures were corrected 
by interpolating the statistical corrections in Table 8 of P12.

Finally, category 14 comprises stars that were unclassified in the KIC and did not yield an 
asteroseismic detection in this study. The characterization of unclassified stars 
is described in detail in the next section.

We note that the literature search was optimized for relatively unevolved stars and hence
did not include certain late stages of evolution such as RR\,Lyrae stars. For more 
accurate stellar properties of these stars we refer the reader to spectroscopic 
follow-up studies that have not been considered here \citep[e.g.,][]{nemec13}.

\begin{table}
\begin{footnotesize}
\begin{center}
\caption{Categories of literature input}
\begin{tabular}{c c c c c}        
\hline         
$C$ & $\teff$ & $\logg$ & [Fe/H] & $N$ \\
\hline
1	& 	spectroscopy	&	asteroseismology	& 	spectroscopy	&	258	\\
2	& 	spectroscopy	&	transits     		& 	spectroscopy	&	20	\\
3	& 	spectroscopy	&	asteroseismology	& 	photometry		&	7	\\
4	& 	photometry		&	asteroseismology	& 	photometry		&	429	\\
5	& 	photometry		&	asteroseismology	& 	KIC				&	12488	\\
6	& 	unclassified	&	asteroseismology	& 	unclassified	&	2762	\\
7	& 	spectroscopy	&	spectroscopy		& 	spectroscopy	&	486	\\
8	& 	spectroscopy	&	photometry			& 	spectroscopy	&	32	\\
9	& 	spectroscopy	&	photometry			& 	photometry		&	310	\\
10	& 	phot./KIC		&	phot./KIC			& 	spectroscopy	&	23	\\
11	& 	photometry		&	photometry			& 	photometry		&	3904	\\
12	& 	photometry		&	KIC					& 	KIC				&	135278	\\
13	& 	KIC				&	KIC					& 	KIC				&	32042	\\
14	& 	unclassified	&	unclassified		& 	unclassified	&	8429	\\
\hline
All	& 	---			&	---				& 	---			&	196468	\\
\hline
\end{tabular} 
\label{tab:priority} 
\end{center}
\flushleft Note: $C$ indicates the priority for each category, and $N$ denotes the number 
of stars in that category.
\end{footnotesize}
\end{table}

\section{Unclassified Stars}
\label{sec:uncl}

Approximately 7\% of \kep\ targets do not have stellar properties 
listed in the KIC due to the lack of data in one or more filters. 
As described in \citet{batalha10} these targets are 
bright stars ($Kp<14$) which were added to supplement 
the original exoplanet target list. Approximately one quarter of these stars were 
consequently dropped from the target list based on 
a photometric luminosity classification using Q1 observations.
In previous runs of the \kep\ planet detection pipeline \citep{jenkins10b} 
unclassified stars were assumed to have solar properties, while for planet-candidate 
catalogs typical main-sequence values based on $J-K$ 
colors were used. In the following section we describe a first effort for a comprehensive 
classification of unclassified \kep\ targets based on broadband photometry and 
asteroseismology.

\subsection{Asteroseismic Analysis and Luminosity Classification}

\begin{figure*}
\begin{center}
\resizebox{15cm}{!}{\includegraphics{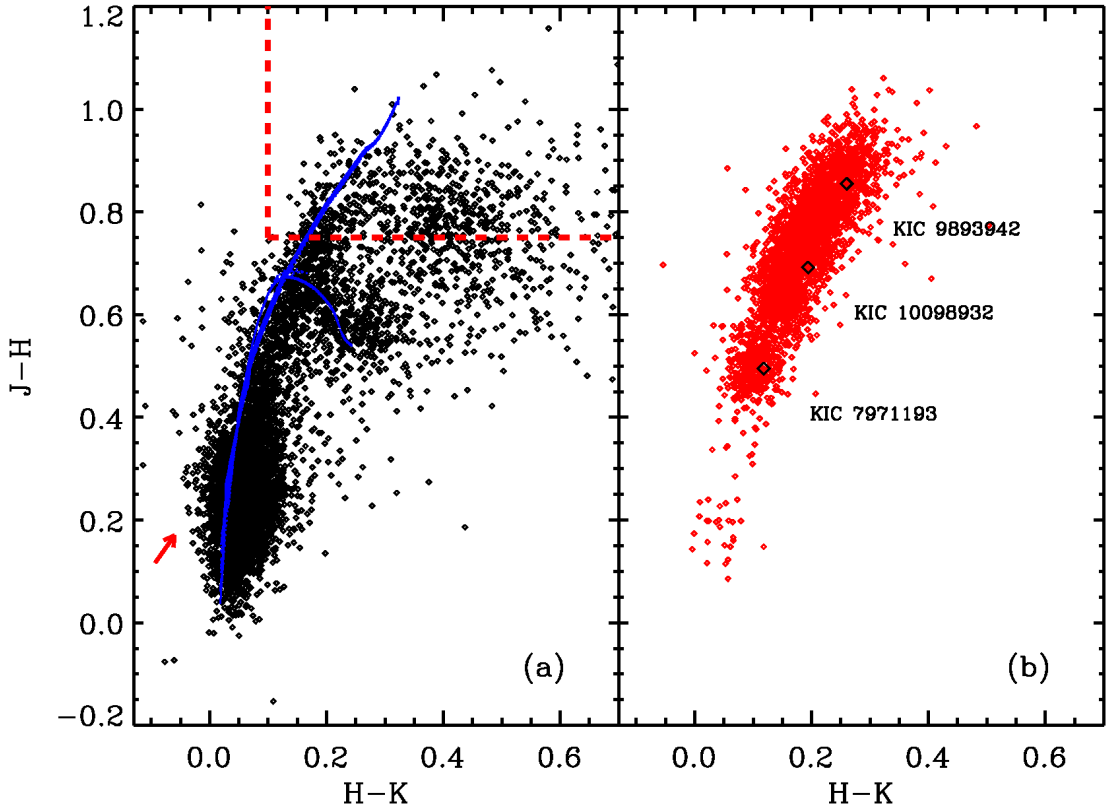}}
\caption{\textit{(a)} 2MASS $J-H$ versus $H-K$ diagram for unclassified stars without 
asteroseismic detections. The blue line shows near-solar metallicity DSEP 
isochrones. The red arrow shows a typical reddening vector for $A_{V}=0.4$\,mag. 
Note that temperature decreases from left to right and from bottom to top, with tracks for 
dwarfs and giant stars separating near $J-H\gtrsim0.7$ and $H-K\gtrsim0.1$. The red-dashed 
box marks the color range for which stars were automatically classified as giant stars 
independent of an asteroseismic detection. 
\textit{(b)} Same as left panel but for unclassified stars with an asteroseismic 
detection. Three examples of oscillating giants are marked, and their power spectra 
are shown in Figure \ref{fig:psexamp}.}
\label{fig:colorcolor}
\end{center}
\end{figure*}

The first step in the classification process was to discern dwarfs from giant stars. 
As described by \citet{batalha10} and \citet{huber13}, asteroseismology is an efficient 
tool to identify giant stars using \kep\ data since the oscillation amplitudes are large 
enough to be detectable independent of shot noise, and because oscillation timescales 
are long enough to be measurable with long-cadence data. We have analyzed 
Q0--14 long-cadence data of 13420 unclassified stars using the 
asteroseismic detection pipelines described by \citet{huber09}, \citet{hekker10c}, 
\citet{mosser09} and \citet{mathur10}.  
We have used simple-aperture photometry (SAP) data as opposed to the Pre-search Data 
Conditioning (PDC) data \citep{smith12,stumpe12} in order to preserve long-periodic oscillations 
typical for high-luminosity giant stars. Instrumental trends such as 
inter-quarter flux discontinuities and pixel-sensitivity dropouts were corrected by 
fitting linear functions to the start and end of each subset, and by applying a 
quadratic Savitzky-Golay filter \citep{savitzky64} with a width of 20 days.
In a few stars, we have also applied the procedures described in \citet{garcia11} to double 
check that features in the light curves were not a consequence of the correction 
procedures and hence verify the reliability of the seismic solutions.

Of the 13420 stars analyzed, 3114 targets showed oscillations that 
classified them as 
giants. A subset of 1760 of these targets have been observed for more than 9 quarters
between 
Q1-16, with 1176 targets of this subset having been observed continuously for the entire 
mission. The new detections presented here raise the  
number of oscillating giant stars detected by \kep\ by $\sim20\%$ 
over previous detections \citep{hekker11c,stello13} to a total of $\sim 15500$ stars. 
Importantly, the newly detected oscillating giants predominantly have surface 
gravities well below the red clump
 ($\logg < 2$, see Section \ref{sec:uncllogg}). Such stars are 
underrepresented in previous asteroseismic samples and provide the opportunity to 
study oscillations in late stages of stellar evolution such as the tip of the red 
giant branch and the asymptotic giant branch \citep{banyai13,mosser13b}. 
Our analysis also yielded 475 classical pulsators, such as $\gamma$\,Doradus 
and $\delta$\,Scuti stars, which were identified using the automated classification 
pipeline by \citet{debosscher11}.

Figure \ref{fig:colorcolor} shows 2MASS $J-H$ versus $H-K$ diagrams for the 
unclassified sample considered in this study. Note that we have rejected stars 
without AAA-quality 2MASS photometry from the catalog, 
reducing the initial sample to 11191 stars (2762 with asteroseismic detections).
Figure \ref{fig:colorcolor}a shows the unclassified 
sample without asteroseismic detections. The blue lines 
show near-solar metallicity DSEP isochrones described in 
Section \ref{sec:models}.
The observations are offset from the models due to reddening (see red arrow in 
Figure \ref{fig:colorcolor}), illustrating the importance of extinction 
when deriving temperatures for unclassified \kep\ stars. 
Note that dwarfs and giants separate 
for $H-K\gtrsim0.1$ (corresponding roughly to spectral type K5), 
with cooler dwarfs retaining a roughly constant $J-H$ color. 
For hotter stars, however, the colors of dwarfs and giants 
overlap and an independent luminosity classification, e.g. 
from asteroseismology or spectroscopy, is required.

\begin{figure}
\begin{center}
\resizebox{\hsize}{!}{\includegraphics{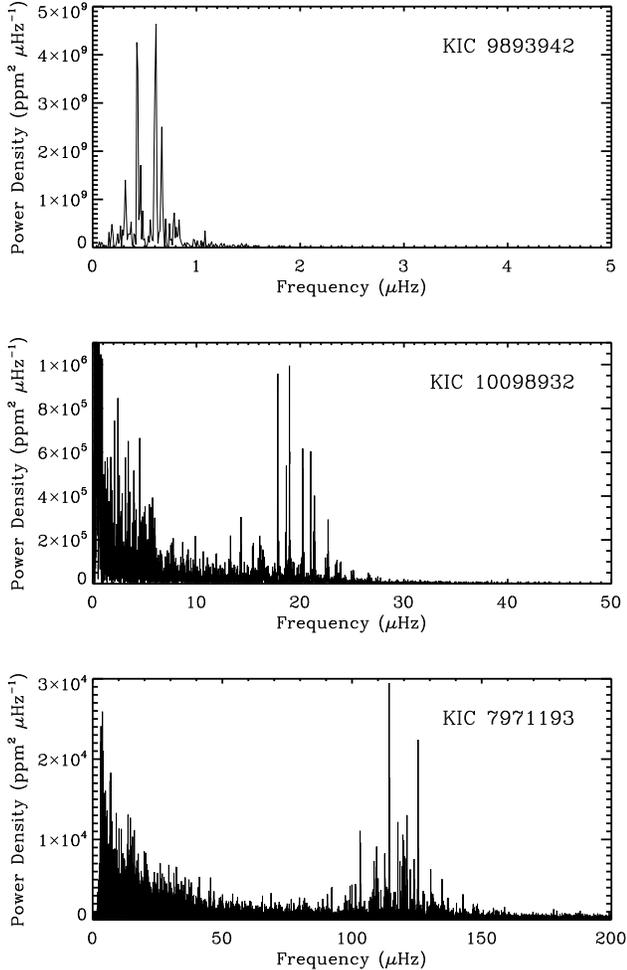}}
\caption{Power spectra of three unclassified stars marked in Figure \ref{fig:colorcolor}b, 
showing clear power excess typical for convection-driven oscillations in giant stars. 
Note that more evolved giants with 
cooler temperatures (larger $J-H$ and $H-K$ in Figure \ref{fig:colorcolor}b) 
oscillate with lower frequencies and larger amplitudes.}
\label{fig:psexamp}
\end{center}
\end{figure}

Figure \ref{fig:colorcolor}b shows the same diagram but for 
unclassified stars with asteroseismic detections.
As expected, asteroseismic detections are mainly found in cool stars in the top 
right section of the plot. 
Three typical examples of oscillating giants are marked, and their power 
spectra showing the presence of convection-driven oscillations are illustrated 
in Figure \ref{fig:psexamp}. Note that as giants become cooler and more luminous 
(larger $J-H$ and $H-K$ colors), their oscillation periods and amplitudes increase.

Notably, there is a significant fraction of stars in Figure \ref{fig:colorcolor}a 
for which an asteroseismic detection 
would have been expected based on their 2MASS colors. Reasons for an asteroseismic 
non-detection in giants include that the star is too cool and too evolved, resulting in 
pulsation periods that are too long for an unambiguous detection with 14 
quarters of long-cadence data. Similarly, many 
unclassified stars have not been observed for the full mission duration, and hence 
the data provide lower frequency resolution.
This discussion implies that a seismic detection is strong evidence 
that a star is a giant, but
a seismic non-detection does not necessarily imply that a star is a dwarf. We therefore 
applied an additional color cut of $H-K>0.1$ and $J-H>0.75$ (see red box in 
Figure \ref{fig:colorcolor}a) to identify giants based on 2MASS colors only. 
Using this procedure, we classify a total of 3302 giants ($\logg<3.5$) and 7889 
subgiant or dwarf stars ($\logg>3.5$) in the unclassified sample.

To test our luminosity classifications, we restricted our sample to the same color-cut 
applied by \citet{mann12} ($Kp-J>2$, $Kp<14$) and
found a total fraction of giant stars (selected using the asteroseismic classifications 
and 2MASS color cuts) of 90\% using all stars, and 92\% using 
only stars with a full set of \kep\ data. This compares reasonably well to the  
giant fraction of $96$\% found by \citet{mann12} based on photometric and 
spectroscopic classifications. The remaining difference could be due to the fact that 
\citet{mann12} considered both classified and unclassified stars, or implies that 
our crude 2MASS color cut to identify giant stars is too conservative.
We also note that, although all seismic detections were checked by eye, it cannot be 
excluded that the unclassified seismic sample includes a small fraction of 
outliers which were erroneously classified as giants (for example due to blends).
Future work including proper motion measurements will enable an 
improved giant-dwarf discrimination for \kep\ targets.

\subsection{Effective Temperatures}

We determined effective temperatures for unclassified stars by comparing observed 
colors to theoretical values of the DSEP model grid described in Section 
\ref{sec:models}. 
As shown in Figure \ref{fig:colorcolor}, interstellar extinction is significant in this 
sample. To ensure consistency with the other \kep\ targets, we adopted the same 
reddening model as applied in the KIC (see Section \ref{sec:kic}). 
For each star, we calculated 
the distance corresponding to each model $J$-band absolute magnitude and the 
apparent $J$-band magnitude, and determined the reddening at the 
distance and galactic latitude of the star. 
Note that we did not restrict reddening to a maximum value. 
This process was repeated until 
convergence in distance was reached. We adopted the reddening law by \citet{cardelli89}, 
with $A_{V}/A_{J}=0.29$. 
For each star, models were restricted to a given metallicity (see next Section).

\begin{figure}
\begin{center}
\resizebox{\hsize}{!}{\includegraphics{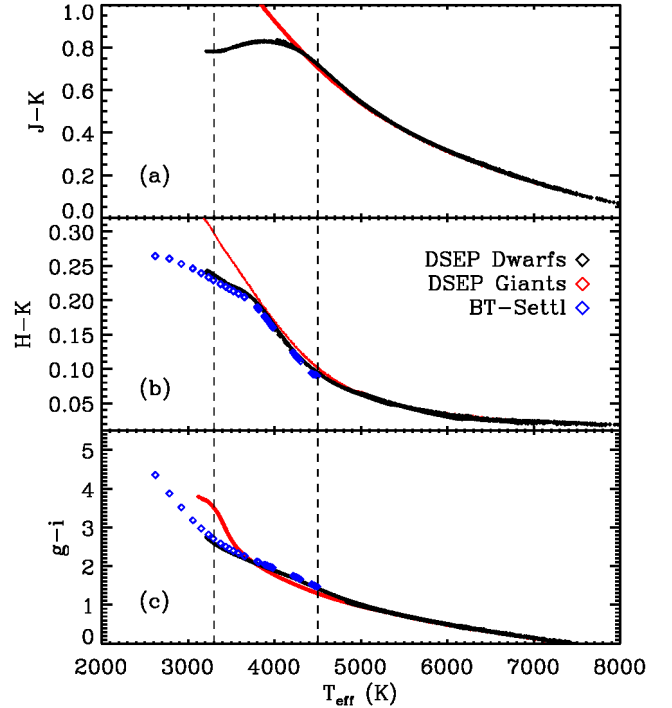}}
\caption{Relations between colors and effective temperatures for near 
solar-metallicity DSEP models for giants (red) and dwarfs (black), as well as 
solar-metallicity BT-Settl models (blue). Dashed lines denote the different 
regimes in which colors were used to derive temperatures: DSEP $J-K$ for stars with 
$\teff\gtrsim4500$\,K, DSEP $H-K$ and $g-i$ for $\teff \sim 3300-4500$\,K, and 
Bt-Settl $H-K$ and $g-i$ for $\teff \lesssim 3300$\,K.}
\label{fig:colorteff}
\end{center}
\end{figure}

The best-fitting model was identified by finding the closest matching model color 
to the observed color. Colors were matched to models depending on the spectral 
type of the star. Figure \ref{fig:colorteff} shows the relation between $J-K$, $H-K$ and 
$g-i$ to effective temperature for DSEP models of dwarfs (black) and giants (red). 
$J-K$ provides the best thermometer for warmer stars with $\teff>4500$\,K. For 
cooler dwarfs, however, $J-K$ becomes insensitive to \teff. $H-K$ shows 
some sensitivity, but only over a color span of $\sim0.1$\,mag, which is relatively 
small compared to typical errors of 0.03\,mag in 2MASS colors. We therefore adopted a 
third color based on $g-i$ from the Kepler-INT survey \citep{greiss12}. 
Note that the Kepler-INT Sloan photometry is in the Vega system, 
which we converted into the AB system using the transformations by \citet{gonzalez11}. 
For stars 
between $3300-4500$\,K, temperatures were then derived from $g-i$ when available, and 
otherwise from $H-K$. Finally, for the coolest dwarfs ($<3300$\,K), 
we applied average polynomial relations from BT-Settl models, as 
described in Section \ref{sec:models}.

\begin{figure}
\begin{center}
\resizebox{\hsize}{!}{\includegraphics{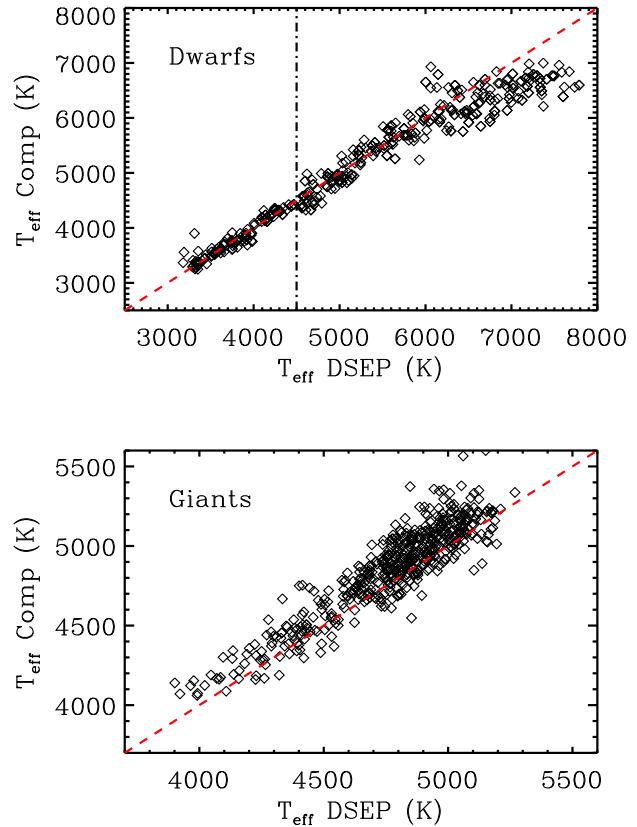}}
\caption{\textit{(a)} Comparison of temperatures
from \citet{dressing13} ($\teff<4500$\,K) and P12 ($\teff>4500$\,K) to 
temperatures derived from fitting single colors to DSEP models for dwarf stars. The red 
dashed line shows the 1:1 relation. 
\textit{(b)} Same as panel (a) but for giant stars. The comparison sample is taken from 
P12.}
\label{fig:compteff}
\end{center}
\end{figure}

To test the derived effective temperatures, we applied the 
same procedure to a subset of stars with available temperatures from 
\citet{dressing13} for dwarfs with $\teff<4500$\,K and P12 for 
dwarfs with $\teff>4500$\,K. The result of this 
comparison is shown in Figure \ref{fig:compteff}a. For stars with $\teff<4500$\,K the 
agreement between the temperatures is excellent, implying that 
Kepler-INT and KIC Sloan colors agree well for these stars. 
For hotter stars the agreement is good for $\teff\lesssim5800$\,K, but we observe 
an increasing systematic offset for hotter stars with DSEP temperatures being 
$\sim$\,500\,K hotter than P12 temperatures at $\teff\sim6500$\,K. 
This offset is caused by the fact that a single color ($J-K$, $H-K$, or $g-i$) 
does not contain independent information on temperature and reddening (or \logg), 
introducing a bias towards hotter, more luminous stars evolving off the main sequence. 
Since temperature and reddening are degenerate, more luminous and distant stars 
with greater reddening and lower \logg\ tend to be selected.

To correct for this bias and ensure consistency of the unclassified stars with the 
remaining sample, we apply an ad-hoc correction for hot stars with the same form as 
adopted by P12:

\begin{equation}
T_{\rm eff,cor} = 5800\,\rm{K}+0.6 (\teff-5800\,\rm{K}) \: ,
\end{equation}

\noindent
for all stars with $\teff>5800$\,K. 

Figure \ref{fig:compteff}b shows a comparison of DSEP temperatures with 
temperatures from P12 for a sample of giant stars. 
We observe an offset, 
with DSEP temperatures being significantly ($\sim 80$\,K) cooler than the 
comparison values. This 
is consistent with the offset between temperatures derived from the 
$J-K$ infrared flux method and temperatures based on Sloan colors discussed by P12. 
Since the offset is generally within the adopted uncertainties, we did not apply a 
systematic correction to temperatures of unclassified giant stars.

\subsection{Surface Gravities and Metallicities}
\label{sec:uncllogg}

The determination of surface gravities and metallicities from broadband colors  
is a notoriously ill-posed problem due to strong degeneracies between fitted 
parameters. Our attempts to perform direct fits of 
observed colors to model colors quickly showed that many stars would be frequently 
matched with physical parameters that are unlikely to occur, such as massive subgiants 
in short-lived phases of stellar evolution. 

\begin{figure}
\begin{center}
\resizebox{\hsize}{!}{\includegraphics{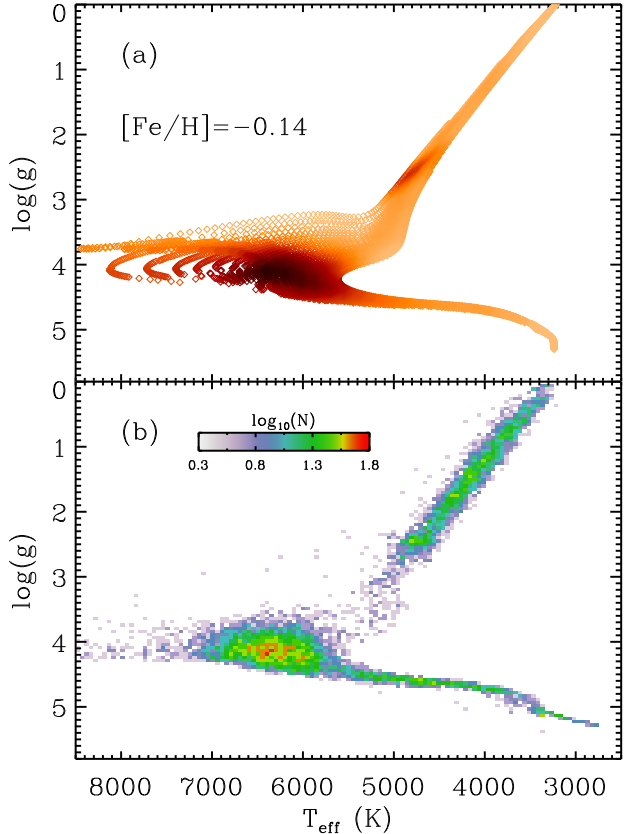}}
\caption{\textit{(a)} Surface gravity versus effective temperature for DSEP models with 
a typical solar-neighborhood metallicity. Each model is color-coded according to the 
prior probability of observing a star in this parameter space, based on a distribution of 
stars in the Hipparcos catalog. Darker colors correspond to a higher prior probability. 
\textit{(b)} Surface gravity versus effective temperature determined for the 
unclassified sample. Colors denote the relative logarithmic 
number density of stars as given in the legend.}
\label{fig:hipprior}
\end{center}
\end{figure}

Following B11, we therefore adopted priors on both surface 
gravity and metallicity. 
The \logg\ prior was constructed using stars in the Hipparcos catalog \citep{vanleeuwen07b} 
with distances $<100$\,pc and fractional parallax uncertainties $<10\%$. 
We first calculated a grid with a stepsize of 0.05\,mag in 
$(B-V)$ and 0.25\,mag in $M_{V}$ (calculated from the parallax assuming $A_{V}=0$) 
and counted the number of Hipparcos stars in each grid cell. 
For each DSEP model at a given metallicity, we then found the grid cell containing the 
$(B-V)$ color and $M_{V}$ magnitude of that model and assigned a prior probability 
corresponding to the number density of Hipparcos stars in that cell. 
Figure \ref{fig:hipprior}a shows the resulting \logg\ prior for a typical solar neighborhood 
metallicity. For the \feh\ prior we adopted the same analytic function as used for the KIC, 
which was constructed from metallicities in the Geneva-Copenhagen survey \citep{nordstroem}. 
We note that the priors have been purposefully chosen to be very similar to those adopted by 
B11 to minimize biases between the unclassified stars and the remaining sample.

Incorporating these priors still resulted in 
distributions of stars that were unrealistically narrow, being confined to the peak of the 
prior distribution at a fixed temperature. This confirmed that
the adopted colors ($J-K$, $H-K$, $g-i$) have little sensitivity to either \logg\ or \feh.
To arrive at more realistic distributions, 
we calculated for each star a one-dimensional prior probability distribution in bins of 0.01\,dex 
in \logg\ around a slice of 50\,K centered on the \teff\ determined in the previous section. 
We then drew a \logg\ value from this distribution with a probability corresponding to the 
prior value at a given \logg. The same procedure was applied to assign metallicities for a 
given star using the B11 metallicity prior. 
Note that for giant stars with asteroseismic detections, surface gravities were calculated 
from the measured frequency of maximum oscillation power, and hence only metallicities were assigned 
in this manner.

The resulting \logg\ versus \teff\ distribution of the unclassified sample is 
shown in Figure \ref{fig:hipprior}b. By construction, the distribution in \logg\ 
closely follows the prior distribution shown in Figure \ref{fig:hipprior}a.
\textit{We stress that the procedure described in the previous paragraph means that 
\logg\ and/or \feh\ for unclassified stars are only statistically accurate, but are 
drawn from a prior probability on a star-by-star basis. 
The properties of these stars (except for temperatures) are therefore not suitable for 
scientific analyses on a star-by-star basis, 
and we strongly encourage follow-up observations
for stars of particular interest (e.g.\ if planet candidates are detected).}
Future efforts will improve the properties 
for these stars by using proper motions or additional colors that 
contain independent information on surface gravities and metallicities.

Figure \ref{fig:hipprior}b also shows 
that there is a significant 
fraction of ultra cool dwarfs in the unclassified sample, which form 
a narrow band of stars with $\teff<3300$\,K. We recall that this discrete distribution is 
caused by the fact that we adopted typical stellar properties for ultra cool dwarfs 
based on polynomial fits to BT-Settl models (see Section \ref{sec:models}). 
We note that $\sim$\,100 of these stars
were matched to the coolest end of the 
BT-Settl polynomials due to very red $H-K$ colors (see Figure \ref{fig:colorcolor}), 
and were excluded from Figure \ref{fig:hipprior}b to avoid biasing the 
color scale.
It is likely that a fraction of these ultra-cool dwarfs are giants, and we 
emphasize that 
the properties for these stars should be used with caution.

\section{Q1--Q16 Catalog}

\subsection{Uncertainties on Input Values}

The procedures described in Sections \ref{sec:consol} and \ref{sec:uncl}
yielded input values for \teff, \logg\ and \feh\ 
for a total of \nstars\ stars. Prior to fitting these constraints to models, 
uncertainties need to be specified for each input value. 
Since uncertainties quoted in the literature are heterogeneous, we 
adopted typical uncertainties on each parameter depending on the observational method 
with which the parameter was derived. 

Asteroseismic surface gravities have been 
shown to be accurate to at least 0.03\,dex \citep{creevey13,morel12,hekker13}, which we 
adopted as a typical uncertainty independent of the evolutionary state of the star. 
Transit-derived densities have proven to be in good agreement with asteroseismic 
densities \citep[e.g.,][]{nutzman11} 
but are based on an implicit assumption of circular orbits, and we hence 
assigned a slightly more conservative uncertainty of 0.05\,dex. Typical spectroscopic 
uncertainties of 2\% in \teff, 0.15\,dex in \logg, and 0.15\,dex in \feh\ 
are based on the comparison of spectroscopic analyses 
with and without asteroseismic constraints \citep{huber13}. 

\begin{figure}
\begin{center}
\resizebox{\hsize}{!}{\includegraphics{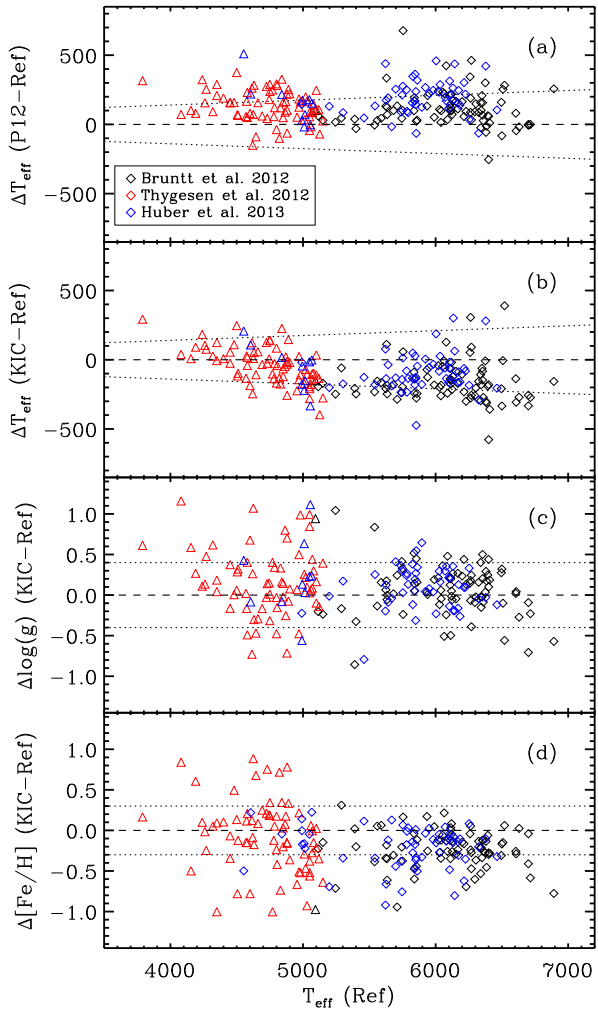}}
\caption{\textit{(a)} Comparison of temperatures by P12 
to temperatures derived for a 
``gold-standard'' sample with constraints from asteroseismology and spectroscopy. 
Dotted lines correspond to the typical uncertainties adopted in this work.
\textit{(b)} Same as panel (a) but for KIC temperatures. 
\textit{(c)} Same as panel (a) but for KIC surface gravities. 
\textit{(d)} Same as panel (a) but for KIC metallicities.}
\label{fig:typicalerrors}
\end{center}
\end{figure}

To estimate typical uncertainties for photometric methods, we have compared 
published results for a sample with combined spectroscopic and 
asteroseismic constraints to properties given in the KIC and P12 
temperatures. The result is shown in Figure \ref{fig:typicalerrors}. 
The median and scatter of the residuals are $+130\pm120$\,K for the P12 
temperatures (panel a), $-130\pm140$\,K for KIC temperatures (panel b), 
$+0.13\pm0.33$\,dex for KIC \logg, and $-0.15\pm0.31$ for KIC \feh.
Based on these residuals and previous 
estimates of uncertainties of KIC properties \citep[e.g.,][]{bruntt11}, we 
assigned uncertainties of 3.5\% in \teff, 
0.4\,dex in \logg, and 0.3\,dex in \feh. We note that the P12 temperatures 
contain homogeneously derived uncertainties based on 
the quality of the input photometry. To preserve this information, 
uncertainties were calculated by adding a 2.5\% systematic 
error in quadrature to the 
formal uncertainty given by P12. 
The typical adopted uncertainties are given in Table \ref{tab:sigma}. 
Table 3 lists the reference 
key for the literature sources of the input values, and Table 4 lists the input 
values with adopted uncertainties for the whole catalog (see Section 6.5 for more details 
on provenances and reference keys).

\begin{table}
\begin{footnotesize}
\begin{center}
\caption{Uncertainties adopted for the input parameters.}
\begin{tabular}{l c c c}        
\hline         
Method & $\sigma_{\teff}$ & $\sigma_{\logg}$ & $\sigma_{\feh}$ \\
	& (\%) & (dex) & (dex) \\
\hline
Asteroseismology	& 	---		&	0.03	& 	---		\\
Transits			& 	---		&	0.05	& 	---		\\
Spectroscopy		& 	2		&	0.15	& 	0.15	\\
Photometry			& 	3.5		&	0.40	& 	0.30	\\
KIC					& 	3.5		&	0.40	& 	0.30	\\
\hline
\end{tabular} 
\label{tab:sigma} 
\end{center}
\flushleft Note: An error floor of 80\,K for spectroscopy and 
100\,K for photometry has been adopted for effective temperatures.
\end{footnotesize}
\end{table}

\subsection{Isochrone Fitting and Derived Uncertainties}
\label{sec:methods}

The input values in Table \ref{tab:input} were fitted to the grid of DSEP 
isochrones to derive radii, masses, densities and luminosities. Matching observations 
to stellar isochrones is a non-trivial task, with
important systematics such as the terminal age bias 
\citep{pont04,jorgensen05,casagrande11,serenelli13}. 
We adopted an approach following \citet{kallinger10} and calculated for each star:

\begin{equation}
\mathcal{L}_{X} = \frac{1}{\sqrt{2\pi}\sigma_{X}} \exp{\left( \frac{-(X_{\rm obs}-X_{\rm model})^2}{2\sigma_{X}^2}\right)}
\end{equation}

\noindent
where $X=\{{T_{\rm eff}, \logg, \rm{[Fe/H]}}\}$ are 
assumed to be independent Gaussian observables. 
The combined likelihood is:

\begin{equation}
\mathcal{L} = \mathcal{L}_{T_{\rm eff}} \mathcal{L}_{\logg} \mathcal{L}_{\rm{Fe/H}} \: .
\label{equ:lh}
\end{equation}

For each parameter, Equation \ref{equ:lh} yields 
a probability distribution that was used to calculate the best-fit, median and 
68\% (1-$\sigma$) intervals. 

The reported value for each stellar property in the catalog corresponds to the 
best-fitting model, which was determined by maximizing Equation (\ref{equ:lh}). 
To specify an error
bar as a single number (as required by the \kep\ planet-detection pipeline), we 
reported the largest distance of the best fit to the upper or lower limit of the 
1-$\sigma$ interval around the median of the probability distribution. 
For highly asymmetric distributions this procedure results in conservative estimates, 
as further discussed below. We have also derived a second set of 
uncertainties by calculating the largest 
difference of the best-fit value to the lower or upper limit of the closest 
1-$\sigma$ interval around the best-fit. 

\begin{figure}
\begin{center}
\resizebox{\hsize}{!}{\includegraphics{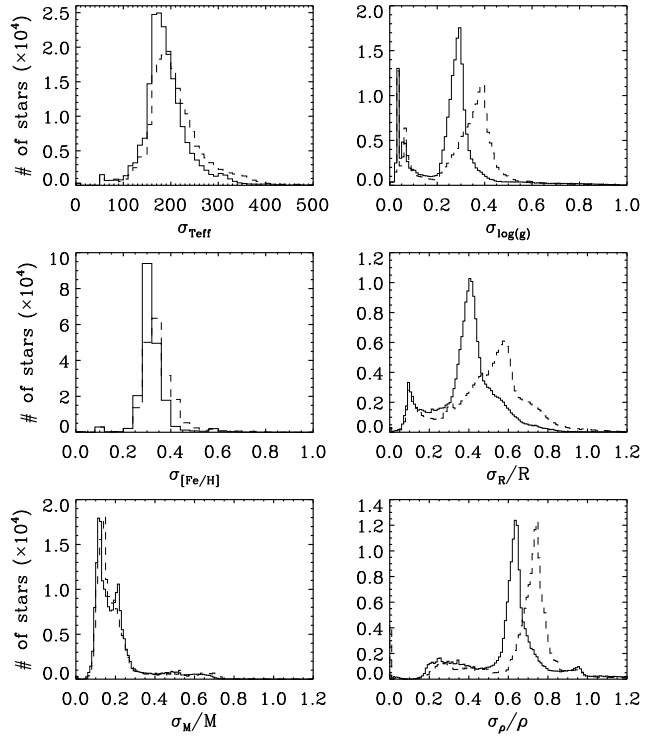}}
\caption{Histograms of uncertainties in \teff, \logg, \feh, $R$, $M$ and $\rho$ for the 
full sample. 
Solid lines show uncertainties based on the 1-$\sigma$ interval closest to the best-fit, 
and dashed lines show uncertainties based on the central 1-$\sigma$ interval (see text).
Note that we show absolute uncertainties for \teff, \logg, and \feh, and 
relative uncertainties for $R$, $M$ and $\rho$.}
\label{fig:sigmahisto}
\end{center}
\end{figure}

Figure \ref{fig:sigmahisto} shows histograms of the derived uncertainties for 
\teff, \logg, \feh, $R$, $M$ and $\rho$. As expected, the uncertainties on 
\teff, \logg, \feh\ largely follow the distribution of uncertainties adopted on the 
input parameters. Uncertainties based on the 
1-$\sigma$ interval around the median (dashed lines) are systematically larger 
than estimates based on the 1-$\sigma$ interval around the best-fit (solid lines)
for higher input uncertainties in \logg, radius and density. This is due to 
asymmetric probability distributions for 
main-sequence stars whose initial \logg\ estimates were based on photometry: assuming 
a 0.4\,dex 
uncertainty, the finite extent of the isochrone grid causes a sharp cut-off at large 
\logg\ values. For smaller fractional uncertainties the distributions become more 
symmetric, and the two uncertainty estimates agree better. Temperature and mass 
mostly yield symmetric distributions, and so this effect does not arise.
Median uncertainties in radius for the full sample over both methods span 
$40-55$\% in radius, and $60-75$\% in density. 
For stellar mass the typical uncertainty is $\sim$20\%, confirming that mass is mainly 
constrained by \teff\ and less affected by large uncertainties in \logg. 
Based on the above discussion, we conclude that the uncertainty estimates based on the 
central 1-$\sigma$ interval (which were used in the \kep\ planet-detection pipeline) 
are probably conservative, especially for main-sequence stars whose input 
surface gravities were based on photometry.

\begin{figure}
\begin{center}
\resizebox{\hsize}{!}{\includegraphics{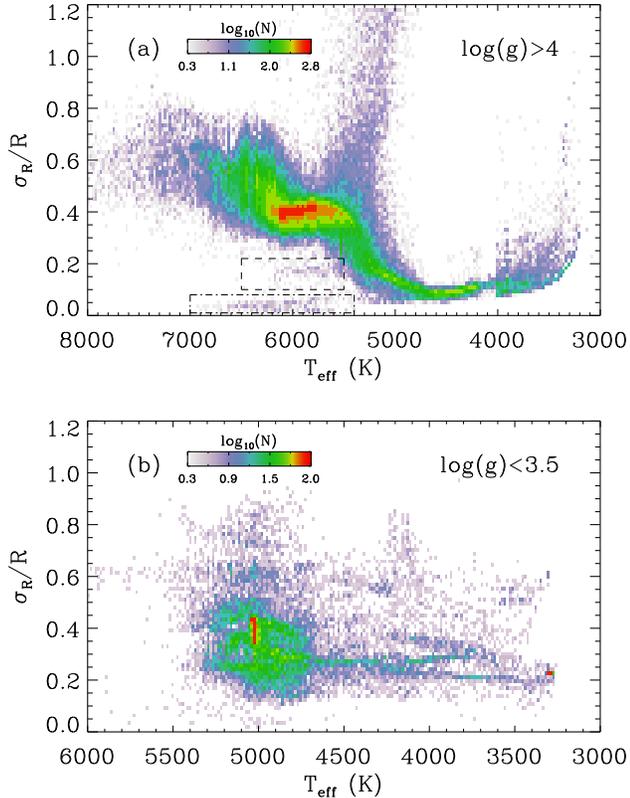}}
\caption{\textit{(a)} Relative radius uncertainty as a function of effective 
temperature for dwarfs $(\logg>4)$. 
Colors denote the relative logarithmic 
number density of stars as given in the legend.
The dashed-dotted and dashed boxes mark stars with surface gravities based on 
asteroseismology and spectroscopy, respectively.
\textit{(b)} Same as panel (a) but for giants $(\logg<3.5)$. 
Note that the majority of these stars have asteroseismic \logg\ measurements.
}
\label{fig:relsigma}
\end{center}
\end{figure}

It is informative to analyze the relative radius uncertainty as a function of 
effective temperature, as shown in Figure \ref{fig:relsigma}a 
\citep[see also][]{gaidos13}. 
Note that we show the uncertainties based on the 
1-$\sigma$ interval around the best-fit here, but the comparison is 
qualitatively similar for the uncertainties based on the 1-$\sigma$ interval around the 
median. For G-type dwarfs 
the median radius uncertainty based on photometry is 40\%, increasing to 
higher values for more massive dwarfs. The ``band'' of points with large uncertainties 
at $\teff\sim5000$\,K is due to K-dwarfs in the 
``No-Man's-Land'' zone (see Section \ref{sec:shortcomings}), 
which have highly bimodal radius distributions 
reaching from the main-sequence to the subgiant branch. For even cooler stars, this 
subgiant degeneracy disappears, and the much slower evolution of stars constrains the radius 
to typical uncertainties of 20\%. We emphasize that these uncertainties do 
not include potential systematic errors in the models, 
which can be significant particularly for cool dwarfs \citep[see, e.g.,][]{boyajian12b}.
Dashed-dotted and dashed boxes highlight the smaller subset 
of stars with constraints from asteroseismology and spectroscopy, respectively. 
As expected, the 
relative uncertainties are smaller, making targets with 
asteroseismic measurements the best characterized stars in the \kep\ field. 

Figure \ref{fig:relsigma}b shows the relative radius uncertainty distribution for 
giant stars. Stars with asteroseismic measurements dominate this sample, with typical 
relative uncertainties in radius of $\sim30\%$. We emphasize that such uncertainties are 
atypically large since we have only used asteroseismic constraints 
on \logg\ and ignored any information on the mean density, which is typically much 
better constrained. Ongoing projects aimed at combining 
APOGEE H-band spectra \citep{eisenstein11} with seismic constraints for Kepler giants 
\citep[the APOKASC project, see][]{meszaros13,apokasc} will
soon provide much improved radii, masses and ages of oscillating giants 
in the Kepler field.

\subsection{Comparisons with Published Radii and Masses}
\label{sec:complit}

\begin{figure}
\begin{center}
\resizebox{\hsize}{!}{\includegraphics{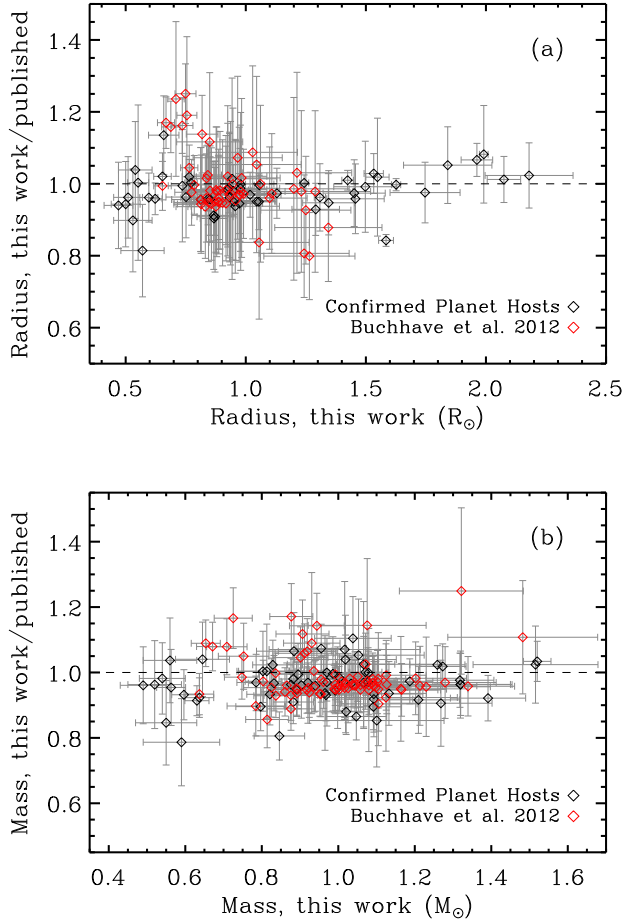}}
\caption{\textit{(a)} Fractional difference between published radii and 
masses and values derived in this work
for confirmed planet host stars (black) and planet-candidate host stars analyzed 
by \citet{buchhave12} (red). 
Only stars with relative uncertainties better than 20\% are shown.
\textit{(b)} Same as panel (a) but for stellar masses.}
\label{fig:compradmass}
\end{center}
\end{figure}

We have compared our catalog results with published radii and masses derived using 
different methods and models. Figure \ref{fig:compradmass} shows a comparison for 
confirmed \kep\ planet host stars taken from the NASA exoplanet 
archive\footnote{http://exoplanetarchive.ipac.caltech.edu/} (black), 
as well as the larger sample of planet-candidate host stars 
by \citet{buchhave12} (red) for stars with relative uncertainties better than 20\%. 
In both cases the majority of the radii and masses 
were derived using Yonsei-Yale (YY) evolutionary tracks \citep{yi01}. 
Overall the residuals show an offset of 1\% with a scatter of 7\% for 
radius and an offset of 3\% with a scatter of 6\% for mass. These offsets, which are 
more pronounced for the sample by \citet{buchhave12}, are likely due 
to differences in the interior models and assumptions of uncertainties on the input values. 
We also observe systematic differences at the low-mass end ($\lesssim 0.8\msun$), 
resulting in a ``kink'' with higher DSEP masses and radii between $\sim 0.6-0.8\msun$, and 
lower DSEP masses and radii for $\lesssim 0.6\msun$. This is consistent 
with systematic differences between DSEP and YY models due to different equations of 
state adopted for low-mass stars \citep{dotter08}. 

\begin{figure}
\begin{center}
\resizebox{\hsize}{!}{\includegraphics{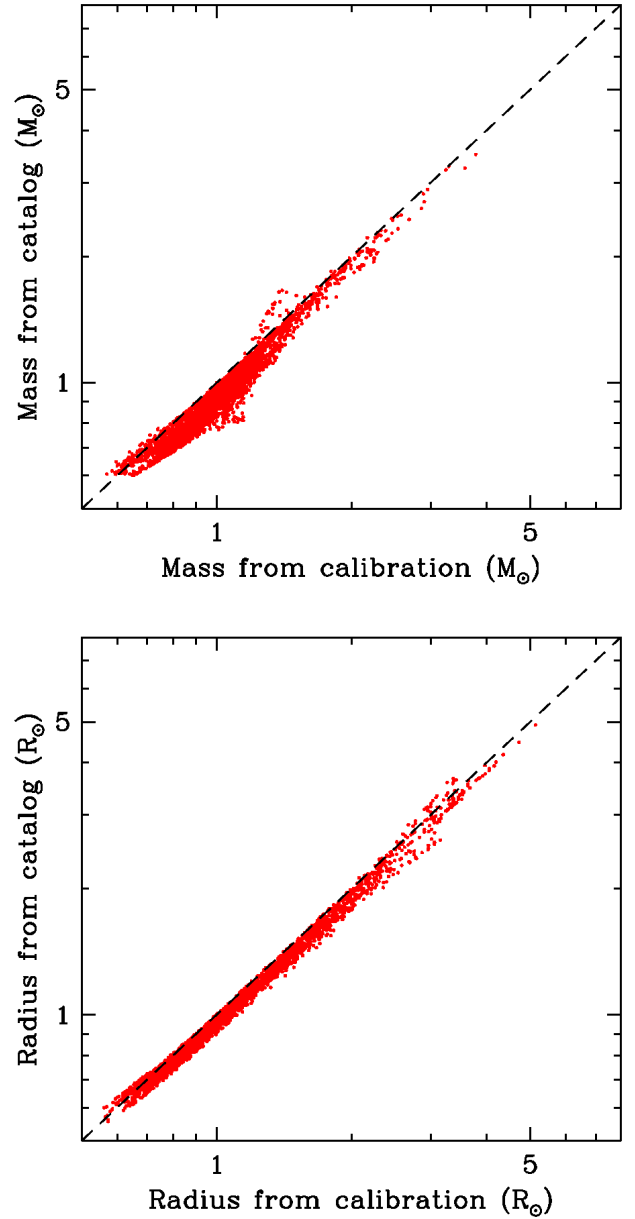}}
\caption{\textit{(a)} Stellar masses in the Q1--Q16 catalog compared to empirical values 
calculated from \teff, \logg\ and \feh\ using the relations by \citet{torres10b}. The 
dashed line shows the 1:1 relation. \textit{(b)} Same as panel (a) but for stellar radii.}
\label{fig:compradmass_torres}
\end{center}
\end{figure}

As an additional test we calculated radii and masses from \teff, \logg\ and \feh\ using
the empirical relations by \citet{torres10b}, which were calibrated using a large 
sample of detached eclipsing binary systems. A comparison with the radii and masses 
derived in this work is shown in Figure \ref{fig:compradmass_torres} for the 
mass range in which the \citet{torres10b} relations are valid. We observe that the 
catalog radii and masses are systematically smaller by $\sim$\,5\% than 
the empirical values. As noted by \citet{torres10b}, a similar offset in mass 
is found when comparing empirically calculated
values to YY isochrone values by \citet{valenti05}, or when using observed values for the 
Sun. Importantly, the offset to the Q1--Q16 catalog does not vary with 
stellar mass and radius, and is typically well within the quoted uncertainties.

\subsection{Catalog Overrides}
\label{sec:overrides}

The primary motivation for fitting \teff, \logg\ and \feh\ to a single set of 
isochrones was to ensure a homogeneous treatment for all stars.
However, methods such as asteroseismology 
provide significantly better constraints on other stellar properties such as the 
mean stellar density. Hence, omitting such additional information can yield significantly 
less accurate stellar radii and masses. More importantly, some of the 
studies used in the consolidation of 
literature values adopted the same models as in this study, hence removing the 
need to re-fit the parameters to ensure consistency with the remaining sample. 
For these reasons, we have adopted literature values for all stars with 
published masses, radii and densities that are based on DSEP models. These studies 
include \citet{dressing13} and \citet{muirhead12} for M dwarfs, and 
\citet{chaplin13} and \citet{huber13} for F-G dwarfs with asteroseismic measurements. 
Note that the latter two studies used a variety of models including 
DSEP, hence providing more robust estimates of uncertainties on stellar 
properties.

\subsection{Final Catalog Description}

The complete Q1--Q16 star properties catalog is presented in 
Table \ref{tab:output}.
For each star we list the best-fitting
\teff, \logg, \feh, radius, mass and density, together with 
the uncertainty based on the 1-$\sigma$ interval around the best fit, 
as described in Section \ref{sec:methods}\footnote{We note that the procedure also 
yielded additional parameters (such as distances) which, however, 
were omitted from this catalog due to the large uncertainties. Additional 
parameters for subsets of stars are available on request, and will be added to 
future updates of the catalog.}. 
For stars with published masses, radii and densities based on DSEP models, stellar 
properties and uncertainties as given in the literature are listed (see Section \ref{sec:overrides}).
Each entry contains 
provenance flags specifying the origin of the input \teff, \logg, and \feh. The 
provenance consists of 
a three letter abbreviation of the method used to derive 
the parameter and a number specifying the reference from which the parameter was 
adopted. The abbreviations are as follows (see also Section \ref{sec:consol}):

\begin{list}{\labelitemi}{\leftmargin=1em}
\itemsep0em 
\item AST = Asteroseismology
\item TRA = Transits
\item SPE = Spectroscopy
\item PHO = Photometry
\item KIC = Kepler Input Catalog
\end{list}

In addition to the provenances for \teff, \logg, \feh, Table \ref{tab:output} also lists a 
provenance 
for the source of interior parameters ($R$, $M$ and $\rho$). The abbreviations are as 
follows:

\begin{list}{\labelitemi}{\leftmargin=1em}
\itemsep0em 
\item DSEP = Derived from the Dartmouth Stellar Evolution Program Models
\item MULT = Derived from multiple evolutionary tracks/isochrones, including DSEP
\end{list}

Note that entries specifying DSEP without a reference number 
correspond to values derived with the model grid 
presented in this work. Interior model flags with reference numbers correspond to 
entries which were replaced by published solutions (see Section \ref{sec:overrides}).

The reference key is provided in Table \ref{tab:refs}. Using the three letter 
abbreviations described above and the reference number, each parameter is directly 
traceable to a single reference and method, and subsets of stars can be filtered 
according to individual methods or references. For example, restricting the sample 
to provenances containing SPE+AST+SPE for \teff+\logg+\feh\ will extract the 
``gold-standard'' sample of stars with combined asteroseismic and spectroscopic 
constraints.

The stellar properties presented in this paper have been adopted in the Q1--Q16 
transit detection run described in \citet{tenenbaum14}. 
We note that the uncertainties 
adopted in that run (and hence displayed in \kep\ pipeline products such 
as data validation reports at the NASA Exoplanet Archive) 
differ from those reported in Table \ref{tab:output} for 
reasons described in Section \ref{sec:methods}. We emphasize, however, that these 
uncertainties 
are not currently used in the \kep\ pipeline, and hence 
this difference does not affect the results in the data validation reports.
We also note that targets that have only been observed in Q0 (commissioning) are not 
analyzed in the transit detection run, and hence have also not been included in this 
catalog. Additionally, $\sim 2000$ stars which are unclassified in the KIC and do not 
have AAA-quality 2MASS photometry remain unclassified in this work, and hence have 
not been included in the catalog (see Section \ref{sec:uncl}).

\section{Comparison to Previous Catalogs}

\begin{figure*}
\begin{center}
\resizebox{15cm}{!}{\includegraphics{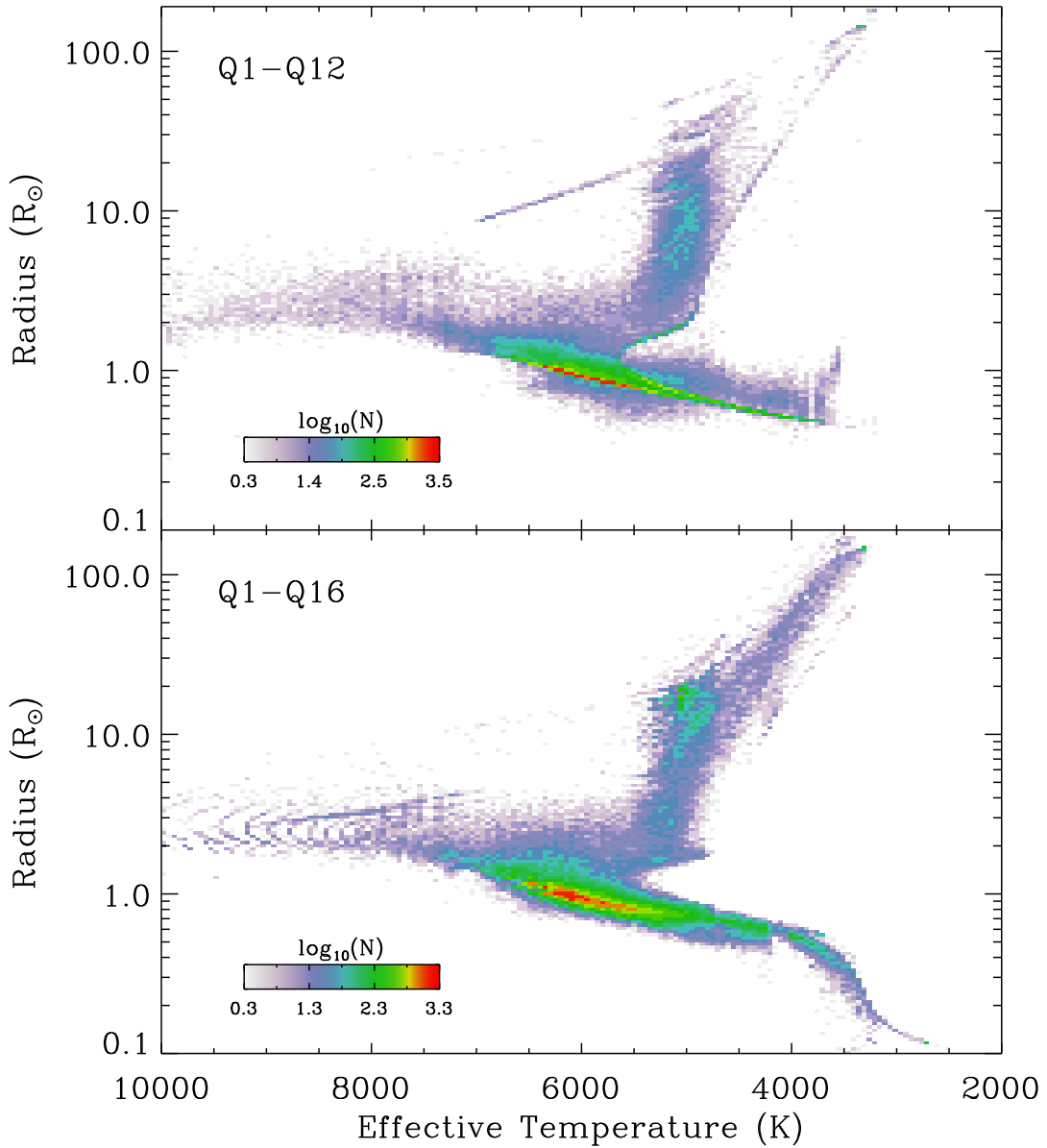}}
\caption{Radius versus temperature for the \kep\ target sample in the Q1--Q12 
star properties catalog (top panel) and the catalog presented here (bottom panel). 
Colors denote the relative logarithmic 
number density of stars as given in the legend.}
\label{fig:compq12}
\end{center}
\end{figure*}

Systematic revisions of stellar properties for \kep\ targets have 
previously been performed for the Q1--Q6 planet-candidate catalog \citep{batalha12}.
Following a similar methodology as 
to this work, constraints from spectroscopic follow-up observations and the KIC 
were fitted to YY models
to reduce well-known biases that are present in the KIC. 
The procedure was subsequently extended to a larger sample of \kep\ targets and the revised 
properties were adopted for the \kep\ transit detection runs producing the 
Q1--Q8 \citep{burke14} and Q1--Q12 \citep{tenenbaum13a,rowe14} planet-candidate catalogs.

Figure \ref{fig:compq12} compares radii as a function of effective temperature for the \kep\ 
target sample as derived in this work to the catalog used 
for the Q1--Q12 transit detection run. Note that we 
compare stellar radii because this is the most important property in the 
context of exoplanet transits.
The Q1--Q12 sample consists of three categories, which are marked with 
different provenances in the NASA Exoplanet Archive: P12 temperatures for a fixed 
metallicity ($\feh=-0.2$) which were combined with KIC \logg\ values and fitted to 
YY tracks (provenance ``Pinsonneault'', 80\% of the sample), 
original KIC values (provenance ``KIC'', 13\% of the sample) and 
unclassified stars which were assumed to have solar values 
(provenance ``Solar'', 7\% of the sample). Note that the latter 
category was excluded from Figure \ref{fig:compq12} to avoid biasing the 
color scale.

Inspection of Figure \ref{fig:compq12} shows several important differences between the 
two catalogs. First, cool M dwarfs now extend to much lower temperatures and radii 
due to the improved coverage of DSEP compared to YY models and the KIC 
in this parameter regime \citep{dressing13}. Second, K dwarfs with
KIC radii $\sim 1\rsun$ as well as G dwarfs with KIC radii $\sim 0.5\rsun$ 
(``No-Man's-Land'' stars) 
are now forced to models compatible with 0.25--15\,Gyr isochrones, resulting in smaller 
radii for the former and larger radii for the latter. 
For K dwarfs this results in a sharp boundary 
corresponding to the oldest isochrone for cool stars. Third, the 
upper red giant branch in the Q1--Q16 catalog is now more populated, with stars reaching up to 
and beyond 100\rsun. The 
majority of these were unclassified stars which have 
now been identified as luminous giants. 

\begin{figure}
\begin{center}
\resizebox{\hsize}{!}{\includegraphics{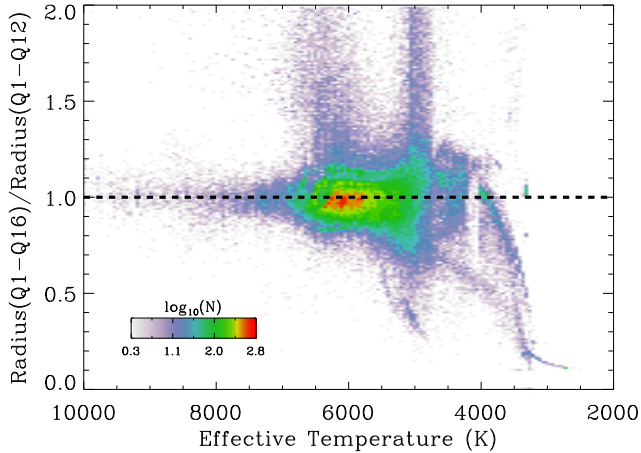}}
\caption{Ratio of radii in the current catalog and the radii in the Q1–Q12 catalog as a 
function of effective temperature. Colors denote the relative logarithmic 
number density of stars as given in the legend.}
\label{fig:compq12_2}
\end{center}
\end{figure}

Figure \ref{fig:compq12_2} shows the ratio between radii presented in this catalog and 
those in the Q1--Q12 catalog as a function of temperature. The median ratio for F--G 
stars is close to one with a scatter of about 10\%. 
This scatter is mostly due to the constant metallicity of $\feh=-0.2$ assumed 
in the Q1--Q12 catalog, whereas for the Q1--Q16 catalog we adopted a
metallicity distribution. 
For certain temperature ranges, large differences in radius can be 
observed. First, the two ``bands'' of stars with radii up to a factor two or more 
larger than in the previous catalog are due to unclassified stars that were 
previously assumed to have solar properties, but have now been classified either as 
giants ($\teff\sim5000$\,K) or stars evolved off the main-sequence ($\teff\sim6000$\,K), 
as well as giants for which KIC \logg\ values were systematically higher than those 
determined from asteroseismology ($\teff\sim5000$\,K). 
Second, K dwarfs ($\sim 4000-5000$\,K) which were peviously fitted to YY isochrones in 
the Q1--Q12 catalog now have radii that are up to $\sim 10$\% larger due to 
model-dependent differences between YY and DSEP models (see Section \ref{sec:complit}).
Third, a large fraction of M dwarfs ($<4000$\,K) have radii that are 
smaller than in the Q1--Q12 catalog. These are predominantly stars 
for which radii were adopted from the original KIC, and have now been 
revised with models more appropriate for cool dwarfs. Remarkably, 
for the coolest dwarfs the updated radii are up to 90\% smaller than those used in the 
Q1--Q12 catalog. 

We emphasize that the large changes in the radii of some Kepler targets will have a 
significant influence on the inferred radii of planet candidates (particularly if the 
host stars were previously unclassified), as well as estimates of planet-occurrence rates. 
The stellar radii presented here should allow 
an improved identification of false-positive planet candidates, as well as the 
identification of interesting candidates orbiting in or near the habitable zones of their 
host stars.

\section{Catalog Shortcomings}
\label{sec:shortcomings}

The methodology in this work inevitably results in a number of shortcomings 
that need to be considered when using this catalog. 

\subsection{General Considerations}

The general trends and biases in the sample can be summarized as follows:

\begin{itemize}

\item For $\sim$\,70\% of all stars the input \logg\ and \feh\ values are still 
based on the KIC, and hence any biases in these values 
\citep[for example potential systematic 
overestimates of \logg\ for G-type dwarfs, see][]{verner11,everett13}
will be included into the Q1--Q16 catalog.  

\item The catalog is based on literature values from a variety of 
techniques, and hence includes systematic offsets between these different methods. 
For example, spectroscopic temperatures are well known to be systematically 
offset from photometric temperatures (see Figure \ref{fig:typicalerrors}).

\item Surface gravities and metallicities for dwarfs and metallicities for giants that 
are unclassified in the KIC are valid in a statistical sense only, but are not 
accurate on a star-to-star basis. Follow-up observations of these 
stars are highly recommended, especially 
if a planet-candidate is detected. Stellar properties of ultra cool dwarfs that were 
previously unclassified in the KIC should also be treated with caution.

\item The adopted isochrone grid does not include He-core burning models for low-mass stars, 
and hence radii, masses and densities for red giant stars are systematically biased as 
they are more frequently matched to higher-mass models which include He-core burning models. 
To derive realistic radii and masses for red giants, it is highly recommended to 
repeat the isochrone or evolutionary track fits using grids which include 
He-core burning models.

\item Uncertainties on stellar properties adopted in the Q1--Q16 transit detection run 
\citep{tenenbaum14} are conservative estimates 
and may be overestimated for cases with very asymmetric probability distributions. 
This particularly applies to F-K dwarfs. The uncertainties presented in 
this paper provide improved estimates which should be unaffected by these biases.
We note, however, that lower or upper 1-$\sigma$ intervals may be significantly 
underestimated for stars near the edge of the model grid.

\item The isochrone fitting method adopted in this study ignores
priors on stellar evolution such as an initial mass function or star formation 
history. Potential biases 
introduced by e.g.\, different evolutionary speeds of stars and different densities of 
models in certain parameter ranges are not yet considered.

\end{itemize}

\subsection{``No-Man's-Land'' Stars}
\label{sec:nomansland}

In their revision of properties of \kep\ planet-candidate host stars, 
\citet{batalha12} identified two groups of stars with KIC 
surface gravities and temperatures that were incompatible with YY isochrones. 
These two groups, namely G-type dwarfs with $\logg \sim 5$ and K dwarfs with 
$\teff \sim 5000$\,K and $\logg \sim 4.2$, were subsequently matched to the closest 
YY isochrone. Lacking any further observational information, 
the fitting procedure in this work results in a similar 
classification of these ``No-Man's-Land'' stars.

\begin{figure}
\begin{center}
\resizebox{\hsize}{!}{\includegraphics{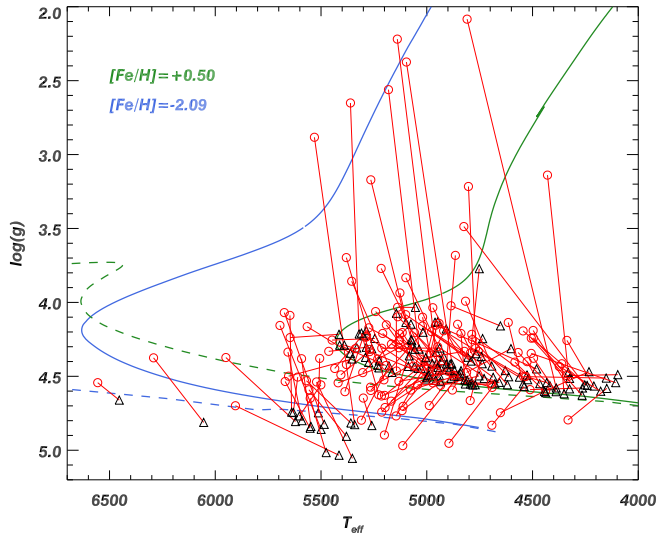}}
\caption{Surface gravity versus effective temperature for a sample of KIC stars in the 
``No-Man's-Land'' zone (black triangles). Solid and dashed lines show 14\,Gyr and 1\,Gyr 
isochrones for two extreme metallicity ranges taken from the BaSTI grid \citep{basti}, which 
were used to select the sample. 
Red circles show the position of the same sample as determined from medium-resolution 
SEGUE spectra. Red lines connect the KIC and SEGUE values for each individual star.}
\label{fig:nomansland}
\end{center}
\end{figure}

To test the accuracy of this procedure, we 
selected stars in the KIC that are either cooler than a 14\,Gyr isochrone with 
$\feh=+0.5$\,dex, or have a \logg\ that is higher than the highest \logg\ of a 1-Gyr 
isochrone with $\feh=-2.09$\,dex. Isochrones for this selection were 
taken from the BaSTI grid \citep{basti}. While these age and metallicity 
cuts are somewhat arbitrary, they do not affect the 
general conclusions presented in this section. The selected sample was then cross-matched to 
the SEGUE catalog of spectroscopic classifications \citep{yanny09}, yielding an 
overlap of 140 stars 
(the total overlap between the KIC and SEGUE includes $\sim$2220 stars, none of which 
are \kep\ targets). 
The comparison of temperatures and surface gravities for this sample 
is shown in Figure \ref{fig:nomansland}. As expected, stars generally 
move away from the ``No-Man's-Land'', with most high-gravity G-type dwarfs moving closer 
to the main-sequence. For the cool ``No-Man's-Land'' sample, stars move in roughly 
equal numbers towards the main-sequence or the subgiant branch. Notably, some of the 
stars are identified as giants in the SEGUE classification. Additionally, a 
considerable number of SEGUE classifications remain in the ``No-Man's-Land'' zone. 
Such targets may correspond to stars with unusual properties or rare evolutionary stages, 
such as merger products, unresolved binary stars, pre-main-sequence stars, or cases in which the 
medium-resolution SEGUE spectra ($R\sim2000$) did not yield a reliable classification.

While the comparison shows that moving stars to 
the nearest isochrone qualitatively yields improved stellar properties, it is 
clear from Figure \ref{fig:nomansland} that for some stars such a procedure can in 
fact yield a \textit{larger} discrepancy to spectroscopic classifications 
than initially given in the KIC. 
We stress that stellar radii for some of these targets may be considerably over- or 
underestimated. 

\section{Conclusions and Future Prospects}
\label{sec:future}

We have presented revised properties for \nstars\ stars observed by the NASA 
\kep\ mission. The main objective of the catalog was to consolidate the large amount 
of stellar characterization work that has been published 
since the launch of \kep\ based on different observation techniques such 
as asteroseismology, spectroscopy, photometry and exoplanet transits. Additionally, we 
estimated the parameters of stars previously unclassified in the KIC, including 2726 
new oscillating red giant stars. The two samples were then combined and homogeneously fit to 
a dense grid of isochrones to derive improved estimates (including uncertainties) 
of temperatures, radii, masses and densities for \kep\ target stars. 
The revised radii and temperatures in the catalog should allow 
an improved identification of false-positive planet candidates and 
planet candidates orbiting in or near the habitable zones of their 
host stars. We emphasize that the present catalog still includes a number 
of important caveats, as summarized in Section \ref{sec:shortcomings}.

Ideally, a catalog of \kep\ targets
should provide the most accurate stellar properties on a star-by-star 
basis, while at the same time being as homogeneous as possible.
The present catalog is somewhat of a compromise between 
accuracy and homogeneity: on one hand, using literature values 
should provide the best possible properties for a given star, while fitting \teff, \logg\ 
and \feh\ to a single 
large grid of isochrones technically ensures that the whole sample is 
internally self-consistent. On the other hand, combining different 
observational techniques inevitably introduces systematic effects that are  
incorporated into the catalog, and are nearly impossible to quantify a posteriori. 
Compared to the KIC the Q1--Q16 catalog is significantly more heterogenous, which 
should be kept in mind for studies vulnerable to biases such as 
planet occurrence rates. However, given the known biases towards unphysical stellar 
properties in the KIC, especially for late-type dwarfs, the approach for the Q1--Q16 
catalog will likely still be an improvement despite the fact that some of the input 
sources are not homogeneous.

Several promising prospects exist to improve further on the current catalog. First, 
it will be essential to provide star properties that are 
independent of KIC-derived properties by refitting the 
broadband photometry to stellar models \citep[e.g.,][]{dressing13,gaidos13b}. 
As partially shown for the unclassified stars in this 
study (see Section 5), the newly available Kepler-INT photometry, which covers nearly 
98\% of all \kep\ targets, holds great promise to complement the available 
KIC $griz$ colors. In particular, the availability of $U$ and $z$ band 
photometry, which is mostly incomplete in the KIC, will be 
essential for improved constraints on reddening, surface gravity and metallicity. 

Improved reddening models will also be essential to derive accurate properties based 
on broadband photometry. The 
APOKASC collaboration \citep{apokasc} has collected H-band spectra for thousands of oscillating 
red giants, which can be combined to derive reddening-independent estimates of 
\teff, \logg\ and \feh. Given the large number of oscillating red giants spread across 
the \kep\ field, this should in principle allow the 
construction of an empirical reddening map. Furthermore, independent 
reddening estimates can be derived using WISE near-infrared photometry \citep{wright10}. 
Large-scale spectroscopic surveys such as APOGEE will also allow measurements of the 
metallicity distribution of stars in the \kep\ field, which may be significantly 
different than the solar neighborhood.

Asteroseismology of 
yet unidentified red giants using long-cadence data will continue to play a major role 
for characterizing \kep\ targets. In particular, a complete census of all 
giants observed by \kep\ will be of prime importance, for example by 
using the detection of oscillations in long-cadence data to identify 
cool stars which might have been misclassified as dwarfs in the 
KIC. Additionally, Figure \ref{fig:categories} shows that a significant number of stars with 
KIC $\logg<3.5$ are not yet included in the asteroseismic sample, potentially indicating 
a significant fraction of misclassified giant stars in the KIC that are subgiants or 
dwarf stars.

One of the most important aspects for future catalogs will be the availability of a 
control sample of stars with well-determined properties. 
Such a control sample will include stars with asteroseismic 
properties, as well as the large number of Kepler Objects of Interest with 
spectroscopic follow-up observations obtained by the Kepler Community Follow-Up 
Program. A particularly 
promising new technique is the determination of empirical surface gravities from the 
measurement of granulation on time scales accessible with \kep\ long-cadence data 
\citep{mathur11b,bastien13}. This new technique has great potential for  
measuring accurate surface gravities for a large number of stars, especially 
if the calibration can 
be extended to a larger parameter space than currently available. Based on this 
potentially large and diverse control sample, 
homogenous transformations from broadband colors to stellar properties for the 
full sample of \kep\ targets can be calibrated, resulting in a catalog which is 
both accurate and homogeneous.

While the \kep\ mission has been a spectacular success for the detection of 
exoplanets and stellar astrophysics in general, 
our understanding of the underlying stellar population of 
the target sample is still limited. Using new and improved techniques and 
follow-up observations, a major future goal will be to improve the characterization of all 
\kep\ targets to maximize the science output both for galactic stellar population studies  
and for studies of exoplanet occurrence rates and populations.

\section*{Acknowledgments}
We thank Lars Buchhave, Bill Cochran, Jonas Debosscher, Joris De Ridder, 
Mathieu Havel, Ulrich Kolb, Dave Latham, Mikkel Lund, 
Phil Muirhead and Angie Wolfgang for helpful 
discussions and comments.
Funding for the \kep\ Mission is provided by NASA's Science Mission Directorate. 
D.H. acknowledges support by an appointment to the NASA Postdoctoral Program at Ames 
Research Center, administered by Oak Ridge Associated Universities through a contract 
with NASA, and support by the Kepler Participating Scientist Program.
S.B.\ ackowledges support from NSF grant AST-1105930 and NASA grant NNX13AE70G.
S.H.\ acknowledges financial support from the Netherlands organisation for Scientific 
Research and ERC starting grant \#338251 (Stellar Ages).
S.M.\ acknowledges support from the NASA grant NNX12AE17G.
F.A.B.\ acknowledges support from a 
NASA Harriet Jenkins Fellowship and a Vanderbilt Provost Graduate Fellowship.
W.J.C.\ acknowledges support from the UK Science and Technology Facilities Council (STFC).
A.M.S.\ is supported  by the MICINN grant AYA2011-24704 and by
the ESF EUROCORES Program EuroGENESIS (MICINN grant EUI2009-04170).
Funding for the Stellar Astrophysics Centre is provided by The Danish National Research 
Foundation (Grant agreement no.: DNRF106). 
The research is supported by the ASTERISK 
project (ASTERoseismic Investigations with SONG and Kepler) funded by the European 
Research Council (Grant agreement no.: 267864).
This publication makes use of data products from the Two Micron All Sky Survey, 
which is a joint project of the University of Massachusetts and the Infrared Processing 
and Analysis Center/California Institute of Technology, funded by the National 
Aeronautics and Space Administration and the National Science Foundation. 
This research has made use of the NASA Exoplanet Archive, which is 
operated by the California Institute of Technology, under contract with the 
National Aeronautics and Space Administration under the Exoplanet Exploration Program.

\bibliographystyle{apj}
\bibliography{/Users/daniel/science/codes/latex/references}

\newpage

\begin{table*}
\begin{footnotesize}
\begin{center}
\caption{Reference Key}
\begin{tabular}{c l l}
\hline
Key & Reference & Methods \\
\hline
 0 & \citet{brown11}			&Photometry				    		\\	    
 1 & \citet{pinsonneault11}		&Photometry				    		\\
 2 & \citet{dressing13}			&Photometry				    		\\
 3 & \citet{buchhave12}			&Spectroscopy  			            \\
 4 & \citet{uytterhoeven11}		&Spectroscopy  			            \\
 5 & \citet{muirhead12}			&Spectroscopy  			            \\
 6 & \citet{bruntt12}			&Spectroscopy/Asteroseismology      \\
 7 & \citet{thygesen12}			&Spectroscopy/Asteroseismology      \\
 8 & \citet{huber13}$^{+}$		&Spectroscopy/Asteroseismology      \\
 9 & \citet{stello13}			&Asteroseismology			    	\\
10 & \citet{chaplin13}			&Asteroseismology			    	\\
11 & \citet{huber11}     		&Asteroseismology			    	\\
12 & \citet{petigura13}			&Spectroscopy  			            \\
13 & \citet{molenda13}			&Spectroscopy  			            \\
14 & \citet{mann12}				&Spectroscopy  			    		\\
15 & \citet{mann13}				&Spectroscopy  			    		\\
16 & \citet{gaidos13b}			&Photometry				    		\\
17 & \citet{martin13}			&Spectroscopy  			            \\
18 & \citet{batalha12}			&Spectroscopy/Transits 			    \\    
19 & \citet{white13}			&Spectroscopy/Asteroseismology		\\	  	  
20 & \citet{bakos10}			&Spectroscopy/Transits/EBs		    \\
21 & \citet{koch10}				&Spectroscopy/Transits/EBs	    	\\
22 & \citet{dunham10}			&Spectroscopy/Transits/EBs		    \\
23 & \citet{jenkins10c}			&Spectroscopy/Transits/EBs		    \\
24 & \citet{holman10}			&Spectroscopy/Transits/EBs		    \\
25 & \citet{lissauer13}			&Spectroscopy/Transits/EBs		    \\
26 & \citet{fortney11}			&Spectroscopy/Transits/EBs		    \\
27 & \citet{endl11}				&Spectroscopy/Transits/EBs	    	\\
28 & \citet{doyle11}			&Spectroscopy/Transits/EBs		    \\
29 & \citet{desert11}			&Spectroscopy/Transits/EBs		    \\
30 & \citet{cochran11}			&Spectroscopy/Transits/EBs		    \\
31 & \citet{ballard11}			&Spectroscopy/Transits/EBs		    \\
32 & \citet{fressin12}			&Spectroscopy/Transits/EBs		    \\
33 & \citet{steffen12}			&Spectroscopy/Transits/EBs		    \\
34 & \citet{fabrycky12}			&Spectroscopy/Transits/EBs		    \\
35 & \citet{lissauer12}			&Spectroscopy/Transits/EBs		    \\
36 & \citet{welsh12}			&Spectroscopy/Transits/EBs		    \\
37 & \citet{orosz12}			&Spectroscopy/Transits/EBs		    \\
38 & \citet{bouchy11}			&Spectroscopy/Transits/EBs		    \\
39 & \citet{santerne11b}		&Spectroscopy/Transits/EBs		    \\
40 & \citet{santerne11}			&Spectroscopy/Transits/EBs		    \\
41 & \citet{muirhead12b}		&Spectroscopy/Transits/EBs		    \\
42 & \citet{bonomo12}			&Spectroscopy/Transits/EBs		    \\
43 & \citet{johnson12}			&Spectroscopy/Transits/EBs		    \\
44 & \citet{nesvorny12}			&Spectroscopy/Transits/EBs		    \\
45 & \citet{orosz12b}			&Spectroscopy/Transits/EBs		    \\
46 & \citet{ballard13}			&Spectroscopy/Transits/EBs		    \\
47 & \citet{meibom13}			&Spectroscopy/Transits/EBs		    \\
48 & \citet{barclay13}			&Spectroscopy/Transits/EBs		    \\
49 & \citet{charpinet12}  		&Spectroscopy/Transits/EBs		    \\
50 & \citet{howell10}			&Spectroscopy/Transits/EBs		    \\
51 & \citet{hebrard13}			&Spectroscopy/Transits/EBs		    \\
52 & \citet{faigler13}			&Spectroscopy/Transits/EBs		    \\
53 & \citet{sanchis13}			&Spectroscopy/Transits/EBs		    \\
54 & This work					&Photometry/Asteroseismology	    \\		
\hline								
\end{tabular}
\label{tab:refs}
\end{center}
\flushleft $^{+}$ Includes references to the following published seismic solutions: 
\citet{barclay12,cd10,batalha11,chaplin12,borucki12,barclay12b,gilliland13,carter12,howell12,huber13b}.
\end{footnotesize}
\end{table*}

\begin{table*}
\begin{footnotesize}
\begin{center}
\caption{Consolidated input values}
\begin{tabular}{c c c c c c c}
\hline
KIC & $\teff$ & $\logg$ & $\feh$ &  P$_{\teff}$ & P$_{\logg}$ & P$_{\feh}$ \\
\hline
  757076  &      $5164\pm154$ &   $3.601\pm0.400$ &  $-0.083\pm0.300$ &  PHO1 &  KIC0 &  KIC0  \\
  757099  &      $5521\pm168$ &   $3.817\pm0.400$ &  $-0.208\pm0.300$ &  PHO1 &  KIC0 &  KIC0  \\
  757137  &      $4751\pm139$ &   $2.378\pm0.030$ &  $-0.079\pm0.300$ &  PHO1 &  AST9 &  KIC0  \\
  757280  &      $6543\pm188$ &   $4.082\pm0.400$ &  $-0.231\pm0.300$ &  PHO1 &  KIC0 &  KIC0  \\
  757450  &      $5330\pm106$ &   $4.500\pm0.050$ &  $-0.070\pm0.150$ & SPE51 & TRA51 & SPE51  \\
  891901  &      $6325\pm186$ &   $4.411\pm0.400$ &  $-0.084\pm0.300$ &  PHO1 &  KIC0 &  KIC0  \\
  891916  &      $5602\pm165$ &   $4.591\pm0.400$ &  $-0.580\pm0.300$ &  PHO1 &  KIC0 &  KIC0  \\
  892010  &      $4834\pm151$ &   $2.163\pm0.030$ &   $0.207\pm0.300$ &  PHO1 &  AST9 &  KIC0  \\
  892107  &      $5086\pm161$ &   $3.355\pm0.400$ &  $-0.085\pm0.300$ &  PHO1 &  KIC0 &  KIC0  \\
  892195  &      $5521\pm184$ &   $3.972\pm0.400$ &  $-0.054\pm0.300$ &  PHO1 &  KIC0 &  KIC0  \\
\ldots & \ldots & \ldots & \ldots & \ldots & \ldots & \ldots \\
 1429653  &      $6636\pm225$ &   $4.622\pm0.400$ &   $0.239\pm0.300$ &  PHO1 &  KIC0 &  KIC0  \\
 1429729  &      $3903\pm136$ &   $4.735\pm0.400$ &  $-0.200\pm0.300$ &  PHO2 &  PHO2 &  PHO2  \\
 1429751  &      $6000\pm185$ &   $4.420\pm0.400$ &  $-0.012\pm0.300$ &  PHO1 &  KIC0 &  KIC0  \\
 1429795  &      $5772\pm164$ &   $4.504\pm0.400$ &  $-0.104\pm0.300$ &  PHO1 &  KIC0 &  KIC0  \\
 1429893  &      $5068\pm143$ &   $4.583\pm0.400$ &  $-0.071\pm0.300$ &  PHO1 &  KIC0 &  KIC0  \\
 1429921  &      $4356\pm125$ &   $4.723\pm0.400$ &  $-0.254\pm0.300$ &  PHO1 &  KIC0 &  KIC0  \\
 1429977  &      $5155\pm180$ &   $4.333\pm0.400$ &  $-0.465\pm0.300$ &  KIC0 &  KIC0 &  KIC0  \\
 1430118  &      $5070\pm155$ &   $3.124\pm0.030$ &  $-0.137\pm0.300$ &  PHO1 &  AST9 &  KIC0  \\
 1430163  &      $6520\pm130$ &   $4.221\pm0.030$ &  $-0.110\pm0.150$ &  SPE6 & AST10 &  SPE6  \\
 1430171  &      $4771\pm166$ &   $4.559\pm0.400$ &  $-0.063\pm0.300$ &  KIC0 &  KIC0 &  KIC0  \\ 
\ldots & \ldots & \ldots & \ldots & \ldots & \ldots & \ldots \\

\hline
\end{tabular}
\label{tab:input}
\end{center}
\flushleft Note that all uncertainties were assigned typical fractional or absolute 
values for a given method, as listed in Table \ref{tab:sigma}. 
Provenance abbreviations: KIC = Kepler Input Catalog, PHO = Photometry, 
SPE = Spectroscopy, AST = Asteroseismology, TRA = Transits. The number at the end of each 
provenance denotes the reference key, as given in Table \ref{tab:refs}. 
(This table is available in its entirety in a machine-readable form in the 
online journal. A portion is shown here for guidance regarding its form and content.)
\end{footnotesize}
\end{table*}

\begin{table*}
\begin{scriptsize}
\begin{center}

\caption{Q1--Q16 Star Properties Catalog}
\begin{tabular}{c | c c c c c c | c c c c}
\hline
KIC & \multicolumn{6}{c|}{Stellar Properties} & \multicolumn{4}{c}{Provenances} \\
 & $\teff$ & $\logg$ & $\feh$ & $R (\rsun)$ & $M (\msun)$ & $\rho$ (g\,cm$^{-3}$) & P$_{\teff}$ & P$_{\logg}$ & P$_{\feh}$ & P$_{M,R,\rho}$ \\
\hline
  757076  &        $5160^{+138}_{-163}$ &   $3.580^{+0.274}_{-0.294}$ &  $-0.100^{+0.260}_{-0.300}$ &      $3.13^{+1.41}_{-1.03}$ &      $1.36^{+0.32}_{-0.40}$ &             $0.062^{+0.112}_{-0.040}$ &  PHO1 &  KIC0 &  KIC0 &  DSEP  \\
  757099  &        $5519^{+183}_{-168}$ &   $3.822^{+0.501}_{-0.276}$ &  $-0.220^{+0.340}_{-0.280}$ &      $2.11^{+1.10}_{-1.02}$ &      $1.08^{+0.33}_{-0.20}$ &                $0.16^{+0.83}_{-0.11}$ &  PHO1 &  KIC0 &  KIC0 &  DSEP  \\
  757137  &         $4706^{+81}_{-103}$ &   $2.374^{+0.029}_{-0.027}$ &  $-0.100^{+0.280}_{-0.340}$ &     $15.45^{+3.57}_{-4.60}$ &      $2.06^{+1.15}_{-1.05}$ &       $0.00079^{+0.00035}_{-0.00014}$ &  PHO1 &  AST9 &  KIC0 &  DSEP  \\
  757280  &        $6543^{+155}_{-206}$ &   $4.082^{+0.228}_{-0.266}$ &  $-0.240^{+0.240}_{-0.300}$ &      $1.64^{+0.82}_{-0.46}$ &      $1.18^{+0.30}_{-0.17}$ &                $0.38^{+0.50}_{-0.24}$ &  PHO1 &  KIC0 &  KIC0 &  DSEP  \\
  757450  &         $5332^{+102}_{-98}$ &   $4.500^{+0.043}_{-0.040}$ &  $-0.080^{+0.160}_{-0.120}$ &   $0.843^{+0.051}_{-0.044}$ &   $0.821^{+0.060}_{-0.040}$ &                $1.93^{+0.30}_{-0.26}$ & SPE51 & TRA51 & SPE51 &  DSEP  \\
  891901  &        $6324^{+153}_{-211}$ &   $4.356^{+0.085}_{-0.327}$ &  $-0.100^{+0.220}_{-0.300}$ &      $1.15^{+0.67}_{-0.14}$ &      $1.08^{+0.28}_{-0.11}$ &                $1.01^{+0.39}_{-0.71}$ &  PHO1 &  KIC0 &  KIC0 &  DSEP  \\
  891916  &        $5602^{+183}_{-148}$ &   $4.587^{+0.039}_{-0.218}$ &  $-0.580^{+0.360}_{-0.280}$ &   $0.741^{+0.286}_{-0.056}$ &   $0.773^{+0.109}_{-0.061}$ &                $2.68^{+0.48}_{-1.50}$ &  PHO1 &  KIC0 &  KIC0 &  DSEP  \\
  892010  &         $4729^{+70}_{-182}$ &   $2.168^{+0.032}_{-0.027}$ &   $0.070^{+0.140}_{-0.470}$ &     $26.09^{+0.44}_{-8.70}$ &   $3.652^{+0.018}_{-2.031}$ &    $0.000290^{+0.000158}_{-0.000018}$ &  PHO1 &  AST9 &  KIC0 &  DSEP  \\
  892107  &        $5080^{+114}_{-155}$ &   $3.354^{+0.261}_{-0.299}$ &  $-0.080^{+0.220}_{-0.320}$ &      $4.29^{+2.02}_{-1.47}$ &      $1.52^{+0.37}_{-0.52}$ &             $0.027^{+0.047}_{-0.018}$ &  PHO1 &  KIC0 &  KIC0 &  DSEP  \\
  892195  &        $5522^{+190}_{-157}$ &   $3.984^{+0.399}_{-0.294}$ &  $-0.060^{+0.280}_{-0.260}$ &      $1.67^{+0.97}_{-0.65}$ &      $0.98^{+0.25}_{-0.10}$ &                $0.30^{+0.93}_{-0.20}$ &  PHO1 &  KIC0 &  KIC0 &  DSEP  \\
 \ldots & \ldots & \ldots & \ldots & \ldots & \ldots & \ldots & \ldots & \ldots & \ldots & \ldots \\
 1429653  &        $6622^{+174}_{-314}$ &   $4.331^{+0.069}_{-0.342}$ &   $0.210^{+0.150}_{-0.430}$ &      $1.32^{+0.76}_{-0.21}$ &      $1.37^{+0.22}_{-0.28}$ &                $0.83^{+0.32}_{-0.59}$ &  PHO1 &  KIC0 &  KIC0 &  DSEP  \\
 1429729  &          $3903^{+76}_{-60}$ &   $4.735^{+0.060}_{-0.070}$ &  $-0.200^{+0.200}_{-0.100}$ &   $0.523^{+0.070}_{-0.050}$ &   $0.541^{+0.070}_{-0.050}$ &                $5.33^{+2.25}_{-2.25}$ &  PHO2 &  PHO2 &  PHO2 & DSEP2  \\
 1429751  &        $6000^{+156}_{-194}$ &   $4.415^{+0.070}_{-0.287}$ &  $-0.020^{+0.220}_{-0.300}$ &      $1.05^{+0.45}_{-0.12}$ &      $1.04^{+0.20}_{-0.12}$ &                $1.27^{+0.40}_{-0.81}$ &  PHO1 &  KIC0 &  KIC0 &  DSEP  \\
 1429795  &        $5768^{+153}_{-150}$ &   $4.501^{+0.045}_{-0.288}$ &  $-0.100^{+0.260}_{-0.280}$ &   $0.906^{+0.377}_{-0.080}$ &   $0.950^{+0.110}_{-0.098}$ &                $1.80^{+0.38}_{-1.14}$ &  PHO1 &  KIC0 &  KIC0 &  DSEP  \\
 1429893  &        $5066^{+158}_{-130}$ &   $4.578^{+0.035}_{-0.090}$ &  $-0.080^{+0.320}_{-0.260}$ &   $0.754^{+0.120}_{-0.059}$ &   $0.785^{+0.097}_{-0.071}$ &                $2.58^{+0.45}_{-0.76}$ &  PHO1 &  KIC0 &  KIC0 &  DSEP  \\
 1429921  &        $4358^{+127}_{-136}$ &   $4.673^{+0.042}_{-0.051}$ &  $-0.260^{+0.320}_{-0.320}$ &   $0.606^{+0.061}_{-0.055}$ &   $0.629^{+0.056}_{-0.063}$ &                $3.98^{+0.84}_{-0.75}$ &  PHO1 &  KIC0 &  KIC0 &  DSEP  \\
 1429977  &        $5166^{+192}_{-158}$ &   $4.554^{+0.071}_{-0.848}$ &  $-0.440^{+0.340}_{-0.260}$ &   $0.728^{+1.287}_{-0.076}$ &   $0.692^{+0.185}_{-0.047}$ &                $2.52^{+0.78}_{-2.39}$ &  KIC0 &  KIC0 &  KIC0 &  DSEP  \\
 1430118  &         $5094^{+82}_{-167}$ &   $3.126^{+0.029}_{-0.030}$ &  $-0.100^{+0.180}_{-0.340}$ &      $6.25^{+0.51}_{-1.61}$ &      $1.90^{+0.26}_{-0.85}$ &          $0.0110^{+0.0039}_{-0.0012}$ &  PHO1 &  AST9 &  KIC0 &  DSEP  \\
 1430163  &          $6520^{+84}_{-84}$ &   $4.221^{+0.013}_{-0.014}$ &  $-0.110^{+0.090}_{-0.090}$ &   $1.480^{+0.030}_{-0.030}$ &   $1.340^{+0.060}_{-0.060}$ &             $0.577^{+0.024}_{-0.025}$ &  SPE6 & AST10 &  SPE6 &MULT10  \\
 1430171  &        $4771^{+185}_{-155}$ &   $4.566^{+0.050}_{-0.051}$ &  $-0.060^{+0.300}_{-0.280}$ &   $0.729^{+0.073}_{-0.067}$ &   $0.714^{+0.092}_{-0.058}$ &                $2.60^{+0.60}_{-0.46}$ &  KIC0 &  KIC0 &  KIC0 &  DSEP  \\
 
\ldots & \ldots & \ldots & \ldots & \ldots & \ldots & \ldots &\ldots & \ldots & \ldots & \ldots  \\
\hline
\end{tabular}
\label{tab:output}
\end{center}
\flushleft Reported are for each parameter the best-fitting value and the lower and upper 
limit of the 68\% interval closest to the best-fit. 
Uncertainties set to zero indicate that no uncertainty estimate is available.
Provenance abbreviations: KIC = Kepler Input Catalog, PHO = Photometry, 
SPE = Spectroscopy, AST = Asteroseismology, TRA = Transits, DSEP = Based on Dartmouth models, 
MULT = Based on multiple models (including DSEP). The number at the end of each 
provenance denotes the reference key, as given in Table \ref{tab:refs}. 
(This table is available in its entirety in a machine-readable form in the 
online journal. A portion is shown here for guidance regarding its form and content. 
An interactive version of this table is available at the NASA Exoplanet Archive: 
http://exoplanetarchive.ipac.caltech.edu/.)
\end{scriptsize}
\end{table*}

\end{document}